\documentclass[twocolumn,letterpaper,pre,superscriptaddress]{revtex4}
\usepackage[T1]{fontenc}
\usepackage[latin9]{inputenc}
\usepackage{amsmath}
\usepackage{graphicx}
\usepackage{amssymb}
\usepackage{hyperref}
\usepackage{color}

\newcommand{\ddt}[1]{\frac{d}{dt}#1}
\newcommand{\E}{[\mathrm{E}]}
\newcommand{\Ss}{[\mathrm{S}]}
\newcommand{\ES}{[\mathrm{ES}]}
\newcommand{\Prod}{[\mathrm{P}]}

\begin{document}

\title{Information topology identifies emergent model classes}

\author{Mark K.~Transtrum}
\affiliation{Department of Physics and Astronomy, Brigham Young University, Provo, Utah 84602, USA}
\email{mktranstrum@byu.edu}

\author{Gus L.~W.~Hart}
\affiliation{Department of Physics and Astronomy, Brigham Young University, Provo, Utah 84602, USA}

\author{Peng Qiu}
\affiliation{Department of Biomedical Engineering, Georgia Tech and Emory University, Atlanta, Georgia 30332, USA}

\begin{abstract}
We develop a language for describing the relationship among observations, mathematical models, and the underlying principles from which they are derived.  Using Information Geometry, we consider geometric properties of statistical models for different observations.  As observations are varied, the model manifold may be stretched, compressed, or even collapsed.  Observations that preserve the structural identifiability of the parameters also preserve certain topological features (such as edges and corners) that characterize the model's underlying physical principles.  We introduce \emph{Information Topology} in analogy with information geometry as characterizing the ``abstract model'' of which statistical models are realizations.  Observations that change the topology, i.e., ``manifold collapse,'' require a modification of the abstract model in order to construct identifiable statistical models.  Often, the essential topological feature is a hierarchical structure of boundaries (faces, edges, corners, etc.) which we represent as a hierarchical graph known as a Hasse diagram.  Low-dimensional elements of this diagram are simple models that describe the dominant behavioral modes, what we call \emph{emergent model classes}.  Observations that preserve the Hasse diagram are diffeomorphically related and form a group, the collection of which form a partially ordered set.  All possible observations have a semi-group structure.  For hierarchical models, we consider how the topology of simple models is embedded in that of larger models.  When emergent model classes are unstable to the introduction of new parameters, we classify the new parameters as relevant.  Conversely, the emergent model classes are stable to the introduction of irrelevant parameters.  In this way, information topology provides a general language for exploring representations of physical systems and their relationships to observations.
\end{abstract}

 
\maketitle

\section{Introduction}
\label{sec:Introduction}

Mathematical modeling is a central component of nearly all scientific inquiry.  Simple models of real systems, coupled with robust methods for interacting with them, is one of the primary engines for scientific progress\cite{wigner1995unreasonable,anderson1972more,machta2013parameter}.  From an information theoretic perspective, mathematical models act as both a type of ``information container'' for storing the answers to experimental questions (in the form of parameter confidence regions, for example) and as a transfer mechanism for using that information to predict the outcome of new experiments.  In order to facilitate these functions, models reflect the properties of the physical system that are relevant for explaining the outcome of experiments while disregarding irrelevant degrees of freedom.  Generically, which properties are relevant or irrelevant depend on the observations, and a complete understanding of the system is best achieved when the relationship among all the potential representations and observations is characterized.  

For example, a complete microscopic description of an ideal gas is neither necessary nor insightful for understanding its thermodynamic properties.  On the other hand, a macroscopic description has a limited domain of applicability and masks the microscopic mechanisms that control the system behavior.  By relating the macroscopic description (temperature and pressure) to microscopic variables (kinetic energy) one simultaneously explains the emergence of the thermodynamic model and identifies its domain of applicability (i.e., the thermodynamic limit).  These deep insights derived from understanding the relationship among different representations and their appropriate domains are lost when either representation is considered alone.

Another context in which this type of insight is useful is that of effective field theories.  Effective field theories demonstrate the utility of including only the appropriate degrees of freedom for describing a particular phenomenon, and the Renormalization Group (RG) makes systematic the process of removing irrelevant degrees of freedom.  Although the actual procedure is usually limited to systems with an emergent scale invariance or conformal symmetry, the analysis makes clear the utility of effective theories and how they emerge from a more complete underlying physical theory.  

Recently it has been suggested that the general success of effective models and emergent theories can be attributed to an information theoretic ``parameter space compression'' resulting from coarse observations\cite{machta2013parameter}.  The present work builds upon this relationship as a vehicle for extending concepts and insights originating in RG to other concepts.   This work should not be interpreted as a direct translation of RG principles, although the concepts we discuss are motivated by analogy with renormalization.

We assume that the appropriate model of a physical system depends not just on the intrinsic properties of the system itself, but also the questions one wishes to ask about it, i.e., the observations to be made at specific experimental conditions.  We develop a new mathematical language, which we call \emph{Information Topology} for describing the relationships among mathematical models, observations, and the underlying principles from which they are derived.  Understanding these relationships leads to a more complete understanding of the physical system.

\begin{figure}
\includegraphics[width=0.75\linewidth]{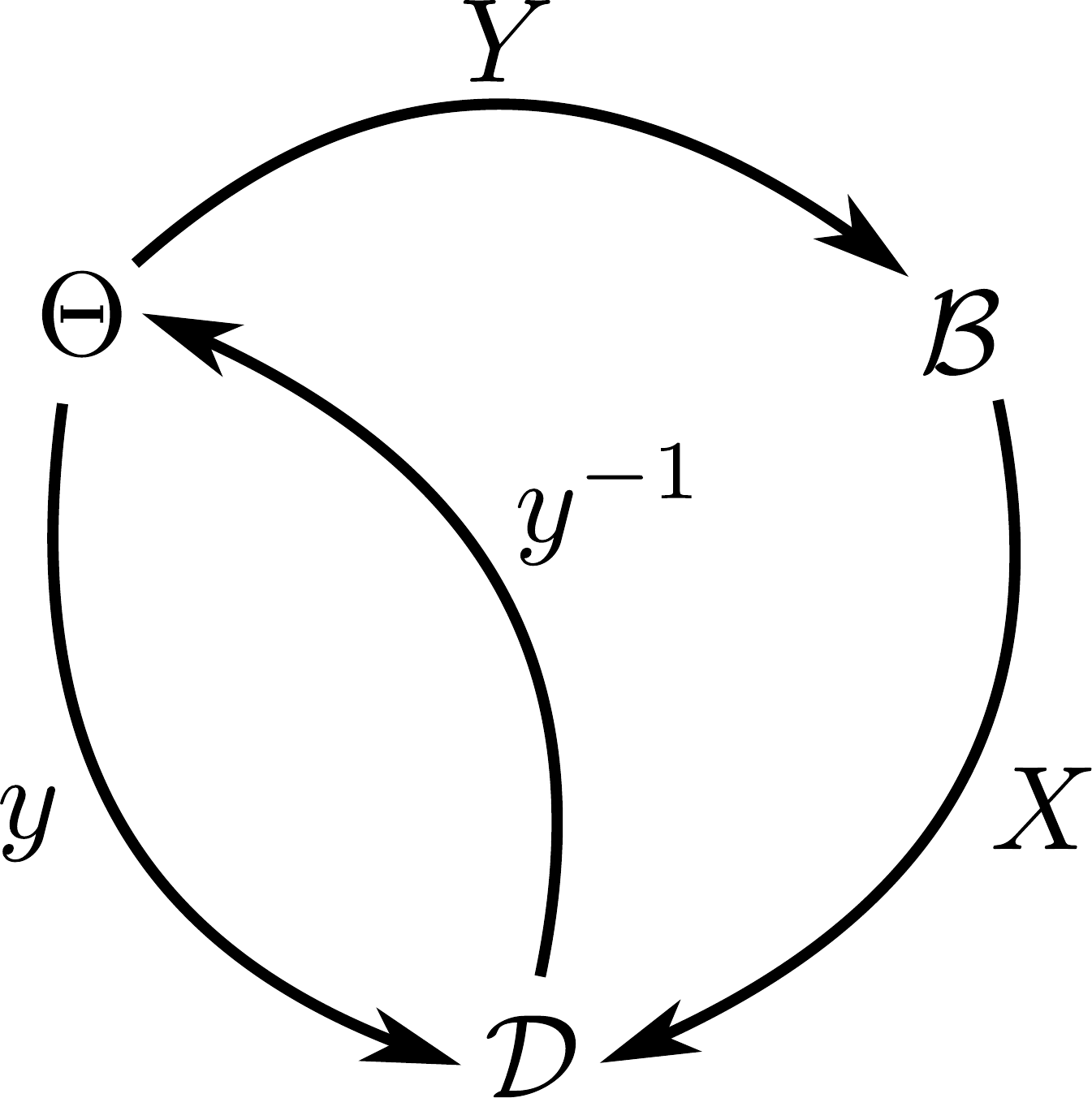}
\caption[]{\label{fig:LeiFig}\textbf{Abstract and Statistical Models} An abstract model, $Y$, maps a parameter space $\Theta$ into a behavior space $\mathcal{B}$.  The behavior space consists of all possible model behaviors under all possible experimental conditions, (even those measurements that may be impractical or impossible).  A real experiment $X$ defines a data space $\mathcal{D}$ as all possible outcomes of the experiment and is a subset of $\mathcal{B}$.  The statistical model, $y$, maps the parameter space into the data space of the specific experiments.  Statistical inference consists of identifying an inverse function $y^{-1}$ that encodes the information from the experiments into the parameter space for extrapolation to new experiments.  Information Topology characterizes the experimental conditions $X$ for which the inverse $y^{-1}$ is well-defined for a given abstract model.}
\end{figure}

The foundation for this theory is a distinction between two types of models as we graphically illustrate in Figure~\ref{fig:LeiFig}.  The first, what we call the \emph{abstract model} and denote by $Y$, is characterized by parameters $\Theta$ and a set of rules that describe the underlying physical principles implemented by the model.  Applying these rules to specific experimental conditions, $X$, leads to what we call the \emph{statistical model} and denote by $y$.  Importantly it is the statistical model that makes falsifiable predictions that can be compared to observations.  We express this relationship by saying a statistical model is a realization of a particular abstract model.  Statistical inference is performed by comparing the statistical model to experimental data through the inverse function $y^{-1}$.  This process induces a metric on the parameter space of the abstract model; however, without any observations, there is no natural metric to the parameter space of the abstract model.

As a concrete example, consider Newton's law of gravitation.  The abstract model is characterized by a parameter (i.e, the universal gravitational constant $G$) and a set of rules (i.e., the inverse square law) for predicting the motion of objects.  Varying the parameter defines a family of abstract gravitational models, and the optimal value is found by comparing the predictions of a statistical model to observation, e.g., predicting the position of the moon, that are a small subset of all possible prediction that the abstract model could make.  The distinguishability of the statistical model's predictions for different values of $G$ defines a metric on the parameter space.  The metric is not intrinsic to the physical system, however, but depends on the observations. Furthermore, the metric would be different if the moon's position were measured to different accuracy or if another experiment (such as observing the position of Mars or measuring the acceleration near the surface of the Earth) were performed.  Each of these potential observations corresponding to different $X$'s in Figure~\ref{fig:LeiFig} and therefore lead to different statistical models $y$.


Since statistical models (unlike abstract models) make predictions for the outcomes of specific experiments, they are naturally interpreted as a manifold of potential predictions embedded in data space.  The interpretation of models as manifolds is the basis for the field of information geometry\cite{rao1949distance,beale1960confidence,bates1980relative,amari1985differential,amari1987differential,kass1989geometry,murray1993differential,amari2007methods,transtrum2010nonlinear,transtrum2011geometry}.  To study the properties of the abstract model, we consider those properties of the model manifold that are invariant to changes in the metric, i.e., what properties are shared by different realizations of the same abstract model.  We therefore consider how geometric properties of statistical models vary depending on specific observations and which properties are invariant to these changes.

\begin{figure}
\includegraphics[width=\linewidth]{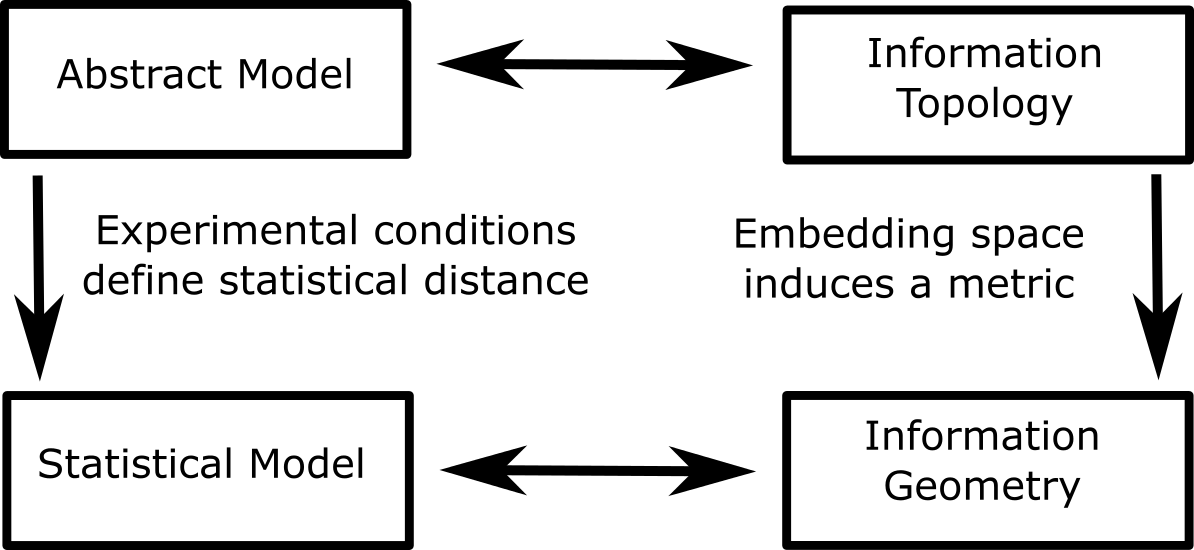}
\caption[]{\label{fig:TopologyGeometry}\textbf{Information Topology and Abstract Models} An abstract model is characterized by a parameter space without a metric.  Comparing the predictions of the model to experimental observations leads to a family of statistical models and imposes a metric on the model's parameter space.  This statistical model can be interpreted as a manifold embedded in the space of all possible experimental outcomes and is the domain of information geometry.  Changing the details of the experimental conditions leads to different models with the same parameter space but different metrics and by extension different model manifolds.  Experimental conditions that do not change the key structural characteristics of the manifold (i.e., are diffeomorphically related to one another) form an information topological description of the abstract model from which they arise.}
\end{figure}

This paper is organized as follows: first, we focus on the observations for which statistical inference can be performed, i.e., for which $y^{-1}$ exists.  We find that the observations for which this holds are diffeomorphisms of the model manifold.  Therefore, properties of the model manifold that are invariant under diffeomorphisms characterize the abstract model.  Information topology is therefore the synthesis of differential topology and information theory just as information geometry is the combination of differential geometry and information theory as we illustrate in Figure~\ref{fig:TopologyGeometry} and elaborate in section~\ref{sec:InformationTopology}.

In section~\ref{sec:collapse}, we consider the effect of coarsening the observations.  Coarsened observations may lead to structural changes in the manifold compared to the identifiable case, what we call \emph{manifold collapse}.  Manifold collapse leads to unidentifiable statistical manifolds that require modifying the abstract models in order to formulate identifiable statistical models. In other cases, coarse observations can lead to a collapse of the boundaries of the model manifold, which in turn leads to practically unidentifiable statistical models.  In many cases, the parameters in a practically unidentifiable model can be arranged in a hierarchy of decreasing relative importance\cite{machta2013parameter}, a phenomenon known as sloppiness and observed in many different models\cite{brown2003statistical,brown2004statistical,frederiksen2004bayesian,waterfall2006sloppy,gutenkunst2007universally,casey2007optimal,daniels2008sloppiness}.  We suggest that sloppiness may be a local manifestation of the collapse of manifold boundaries.

In section \ref{sec:ModelClasses}, we show that the topological structure of an abstract model can be graphically represented as a hierarchical graph known as a Hasse diagram.  This visualization naturally leads to a reinterpretation of the model parameters in terms of refining approximations to simple reduced models\cite{transtrum2014model}.  These reduced models are low dimensional elements of the Hasse diagram and identify distinct behavioral modes of the model.  In this way, the topology of the abstract model identifies the relationships among distinct system behaviors and minimal representations of those behaviors.  We find that set of all possible observations form a semi-group structure with diffeomorphism subgroups characterized by unique Hasse diagrams.

In section \ref{sec:embedding} we consider how topologies of hierarchical models are embedded within one another.  This allows us to identify stable and unstable model classes and classify parameters as either relevant or irrelevant (in analogy with corresponding classifications in the context of renormalization group analysis\cite{goldenfeld1992lectures,zinn2007phase}).

\section{Information Topology and Mathematical Models}
\label{sec:InformationTopology}

To formalize the concepts outlined in the introduction, we consider statistical models that are defined as a probability distribution for specific observations, $P(\xi)$.  Here $P$ represents the probability of observing some outcome $\xi$.    We focus on probability distributions because they have broad applicability and have a natural metric (i.e., the Fisher Information defined below) for measuring distances among model predictions.  This assumption is not strictly necessary, however, and we discuss other possible metrics in the supplement.  We assume these statistical models are derived as specific applications of some abstract model to particular experimental conditions, so that there are many statistical models that share a common parameter space.

We now consider the concept of \emph{identifiability} of a statistical model.  In general, there are two classes of identifiability issues, structural and practical.  A model $y$ with structurally identifiable parameters is an injective map from parameters to predictions so that an inverse, $y^{-1}$ can be constructed.  Structural identifiability is often described as the potential to accurately infer unique parameter values from a model given perfect, noise-free data and is a necessary condition for parameter inference in a real experiment\cite{eisenberg2013identifiability}.  Structural identifiability is further categorized as either global or local.  Global identifiability means that the inferred parameters are unique when considering the entire domain of physically relevant parameters.  Local identifiability, on the other hand, means that there exists an open neighborhood centered on the inferred parameters in which the inferred parameters are unique\cite{rothenberg1971identification,cobelli1980parameter}.

Local identifiability is closely related to the Fisher Information Matrix (FIM):
\begin{equation}
  \label{eq:FIM}
  g = \left\langle \left( \frac{\partial \log P}{\partial \theta} \right)^2 \right\rangle = \left\langle \left( \frac{ \partial^2 \log P}{\partial \theta^2} \right) \right\rangle,
\end{equation}
where $\langle \cdot \rangle$ means expectation with respect to the model.  It can be shown that a model is locally identifiable at parameter values $\theta$ if and only if the FIM is non-singular at $\theta$\cite{rothenberg1971identification}.  Qualitatively, a locally unidentifiable model has redundant parameters so that the parameter values can be changed in a coordinated way without changing the predictions of the model.  The null space of the FIM is precisely the linear subspace in which parameter values can be shifted without changing the model behavior.  

The other type of identifiability is practical.  Although in principle it may be possible to uniquely identify true parameter values, in practice the number of repeated observations necessary to obtain a reasonable estimate may be impractical.  Practical unidentifiability is also related to the FIM through the Cramer-Rao inequality:
\begin{equation}
  \label{eq:CramerRao}
  Cov( \hat{\theta} ) \geq g^{-1}/n,
\end{equation}
where $Cov(\hat{\theta})$ is the asymptotic covariance matrix of an unbiased estimator $\hat{\theta}$ and $n$ is the number of repeated samples.  If the FIM is not singular but approximately so, it may take an unreasonable sample size, i.e., large value of $n$, to obtain accurate estimates.

It is not uncommon, particularly in models with many parameters, for the FIM to be poorly conditioned leading to practical identifiability problems, a phenomenon sometimes known as sloppiness\cite{brown2003statistical,brown2004statistical,frederiksen2004bayesian,waterfall2006sloppy,gutenkunst2007universally,casey2007optimal,daniels2008sloppiness}.  Parameter identifiability can be improved through repeated application of experimental design techniques\cite{faller2003simulation,cho2003experimental,casey2007optimal,balsa2008computational,apgar2008stimulus,apgar2010sloppy,erguler2011practical,chachra2011comment,transtrum2012optimal} or model reduction\cite{transtrum2014model}.

The basic principle behind the geometric interpretation of statistics is that the FIM in Eq.~\eqref{eq:FIM} can be interpreted as a Riemannian metric tensor.  (Although we use the FIM as the measure of distance throughout this paper, our results generalize to any distance measure among models as we discuss in the supplement.)  A parameterized statistical model is therefore equivalent to a Riemannian manifold.  We present here four simple illustrative examples that will then serve to illustrate the topological principles we explore later.

First, we consider a model as the sum and difference of exponentials:
\begin{equation}
  \label{eq:ExpDef}
  \xi_i(\theta_1, \theta_2)  =  \left\{ \begin{array}{l l} 
      e^{-\theta_1 t_i} + e^{-\theta_2 t_i} + z_i & \ 1 \leq i \leq M_s \\

      e^{-\theta_1 t_{i}} - e^{-\theta_2 t_{i}} + z_{i} & \ M_s <  i \leq M_s + M_d
      \end{array} \right.
\end{equation}
where $z_i$ is normally distributed random variable with zero mean and standard deviation $\sigma_i$ and $M_s$ and $M_d$ are the number of observations of the sum and differences respectively.  This model could describe, for example, the radioactivity of a sample with two radioactive components in the case it is possible to distinguish between the two types of radiation, i.e., it is possible to measure both the total radiation and the difference in radiation types as in Eq.~\eqref{eq:ExpDef}.  

\begin{figure}
\includegraphics[width=\linewidth]{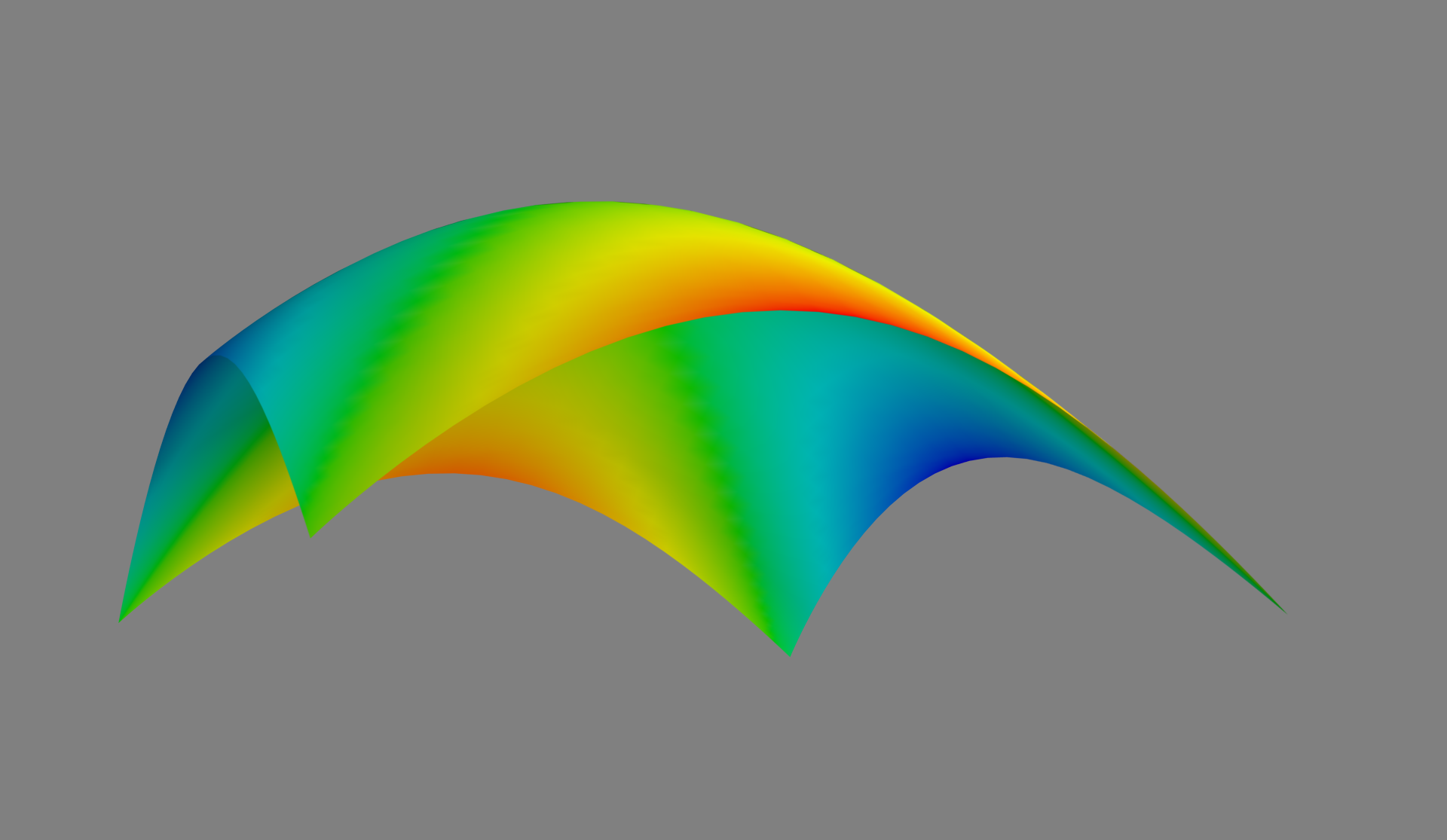}
\caption[]{\label{fig:Exps_1}\textbf{Exponential Model Visualization.} A visualization of the of the model manifold described by Eq.~\eqref{eq:ExpDef} for one possible realization of the abstract model.  An alternative realization is given in Figure~\ref{fig:Exps_2}.}
\end{figure}
\begin{figure}
\includegraphics[width=\linewidth]{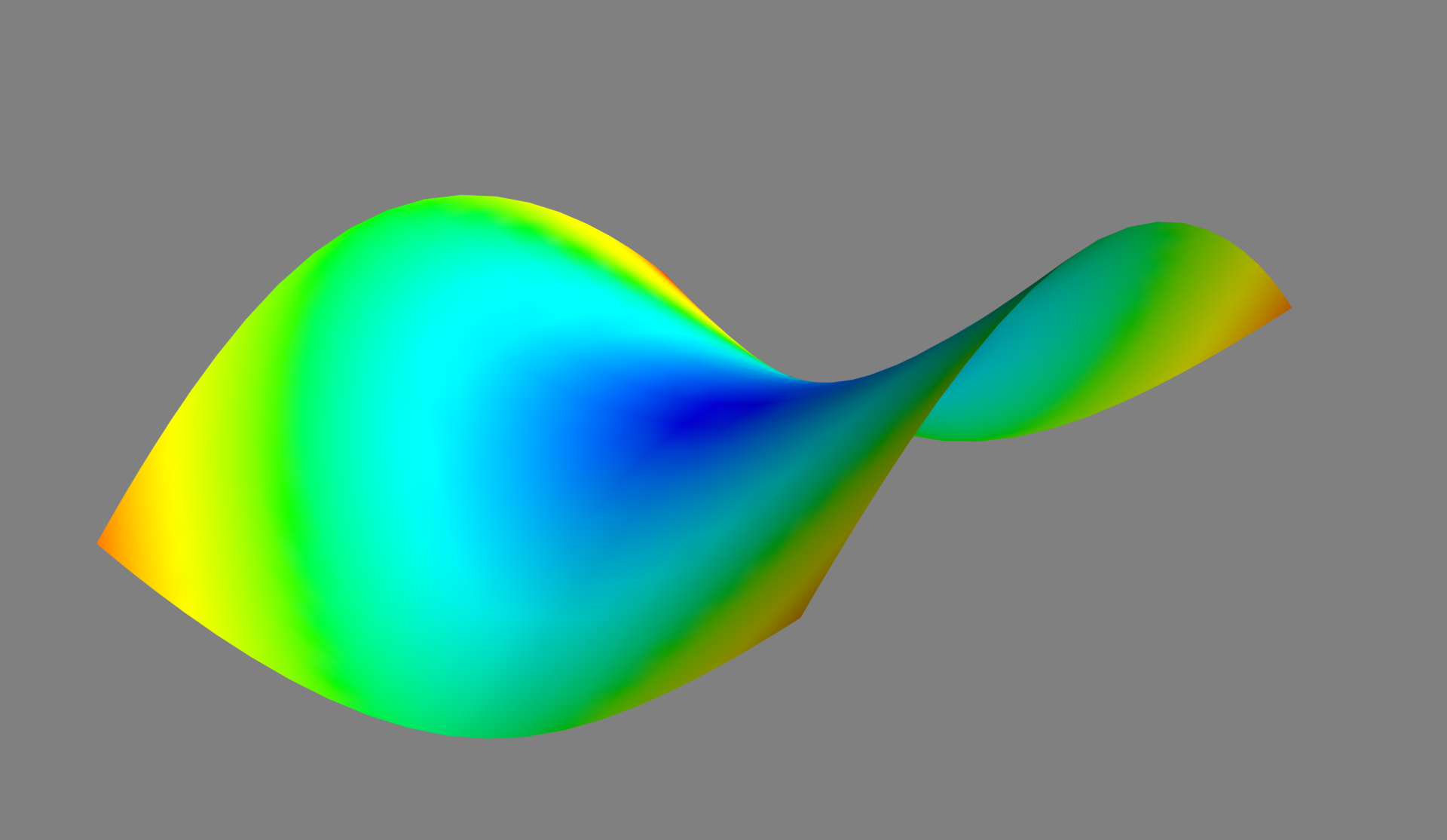}
\caption[]{\label{fig:Exps_2}\textbf{Alternate Exponential Model Visualization.} An alternative realization of the model described by Eq.~\eqref{eq:ExpDef} with different observations that those that led to Figure~\ref{fig:Exps_1}.  Notice that by changing the observations, the geometric properties of the model manifold also change.  The manifold may stretch and bend, but the manifold remains topologically equivalent to a square.}
\end{figure}

For the model in Eq.~\eqref{eq:ExpDef}, model manifolds for two different observations (i.e., different choices of time points $t_i$ and experimental accuracies $\sigma_i$) are given in Figures~\ref{fig:Exps_1} and \ref{fig:Exps_2}.   Specifically, both manifolds correspond to a sampling of time points logarithmically spaced between $0.1$ and $10$.  Figure~\ref{fig:Exps_1} corresponds to $\sigma_i = 1$ for all values of $i$ while Figure~\ref{fig:Exps_2} has $\sigma_i = 1$ for $i < M_s$ and $\sigma_i = 1/2$ for $i > M_s$.  The manifolds are then generated by considering the model predictions for all physically allowed values of $\theta_1$ and $\theta_2$.  Notice that since the observations that led to Figures~\ref{fig:Exps_1} and \ref{fig:Exps_2} are different, geometrical features (e.g., distances and curvatures) of the manifolds are also different.  However, because these manifolds are both identifiable realizations of the same abstract model, they share important structural characteristics.  Specifically in this case, they are both square-like, in the sense of having four edges and four corners.  

Next, we consider two generalizations of the Ising model given by the Hamiltonians
\begin{equation}
  \label{eq:IsingH1}
  H = - \sum_i J_i s_i s_{i+1},
\end{equation}
and
\begin{equation}
  \label{eq:IsingH2}
  H = -\sum_{i,\alpha} J_\alpha s_i s_{i+\alpha}.
\end{equation}
In Eqs.~\eqref{eq:IsingH1} and \eqref{eq:IsingH2}, $s_i$ are spin random variables arranged in a one-dimensional chain that can take values $\pm 1$.  The probability of a particular configuration is then given by a Boltzmann distribution
\begin{equation}
  \label{eq:IsingBoltzmann}
  P \propto e^{-H},
\end{equation}
where we have taken $k_B T = 1$ and the normalization is determined by summing over all configurations.

\begin{figure}
\includegraphics[width=\linewidth]{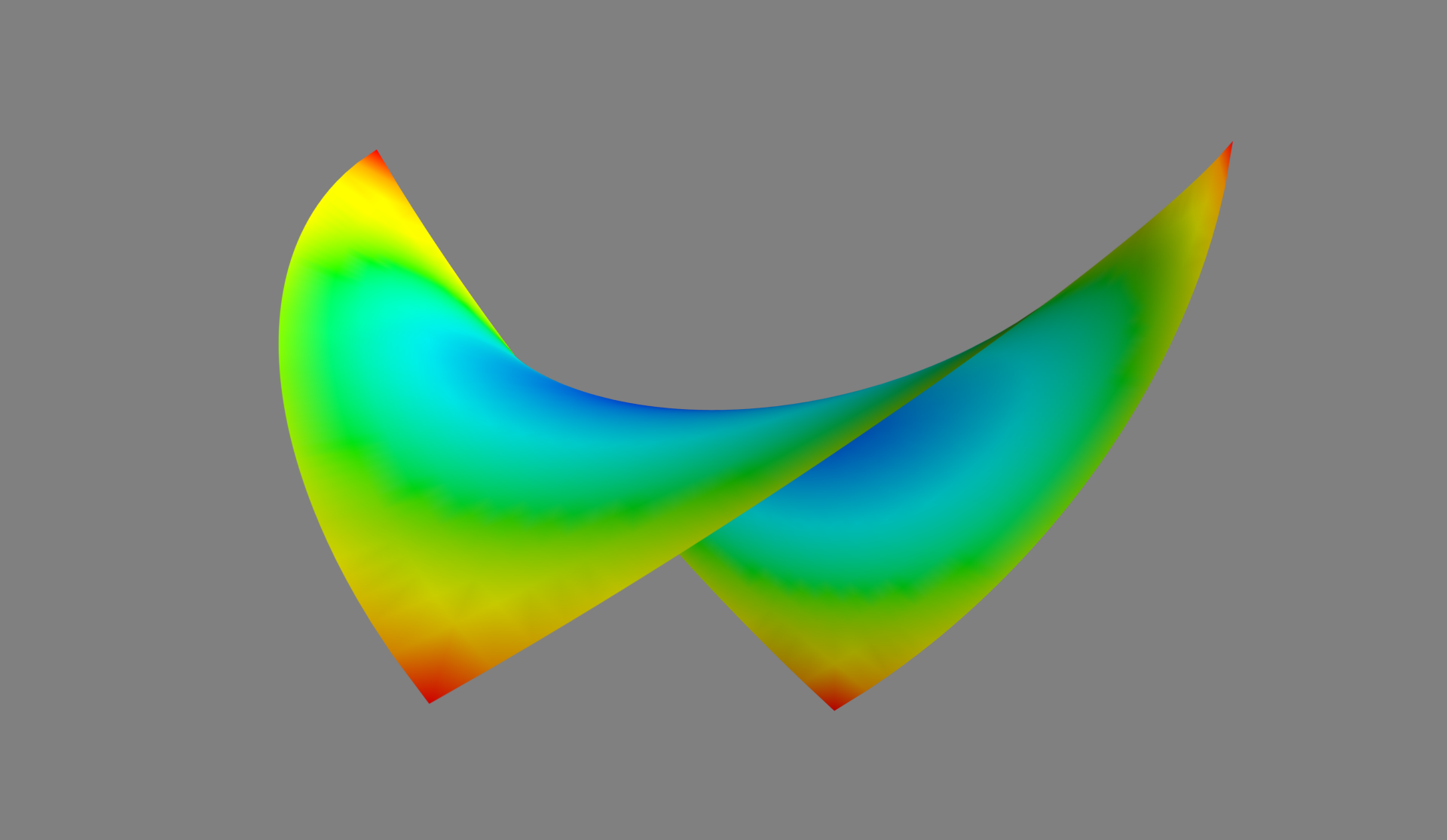}
\caption[]{\label{fig:Ising_3_2_r_1}\textbf{Model Manifold of the Ising Model in Eq.~\eqref{eq:IsingH1}.} The generalized Ising model given by Eq.~\eqref{eq:IsingH1} with three spins and two parameters is topologically a square.}
\end{figure}

\begin{figure}
\includegraphics[width=\linewidth]{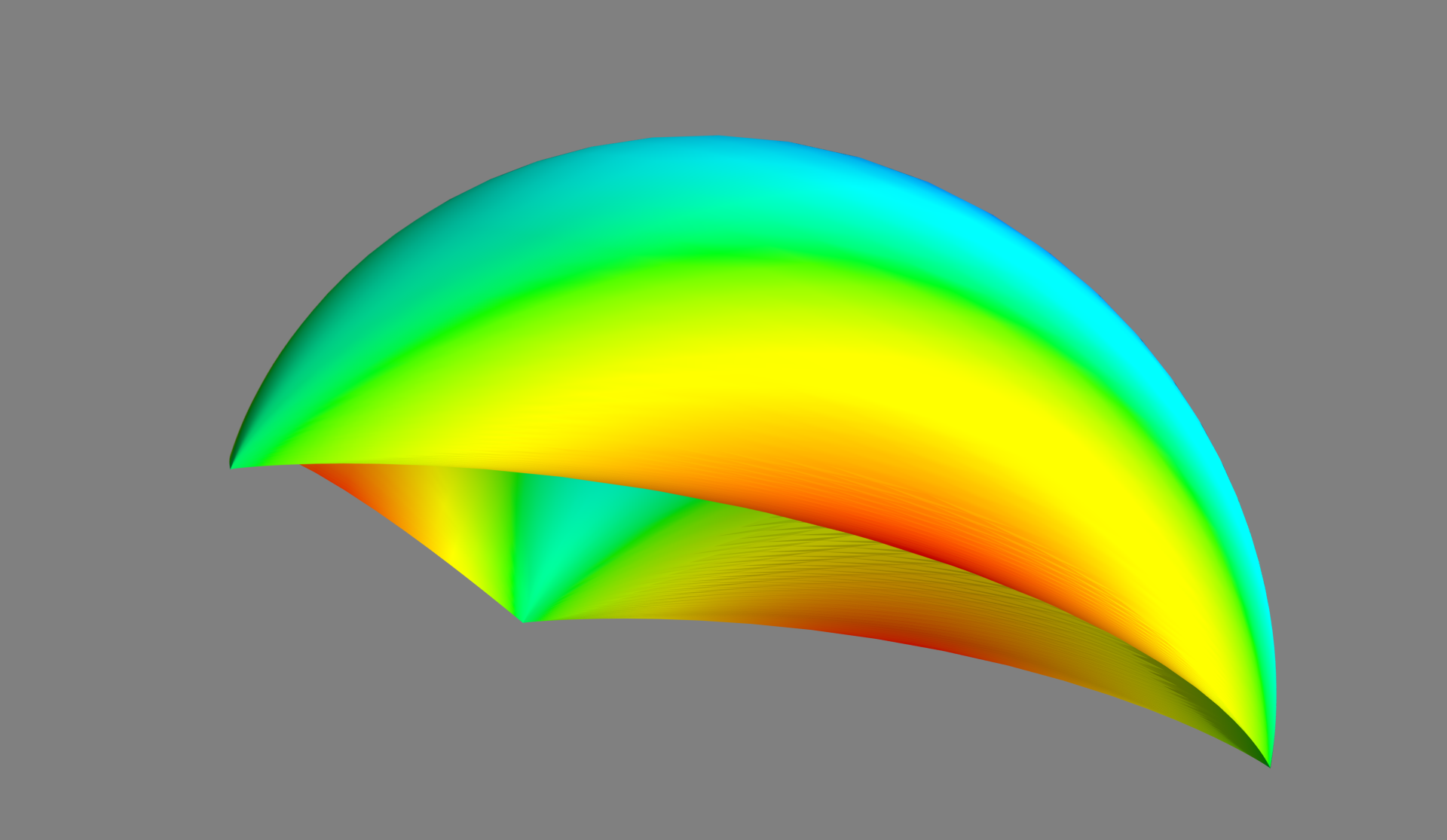}
\caption[]{\label{fig:Ising_4_2_k_1}\textbf{Model Manifold of the Ising Model in Eq.~\eqref{eq:IsingH2}.} The generalized Ising model given by Eq.~\eqref{eq:IsingH2} with four spins and two parameters is topologically a triangle.}
\end{figure}

The parameters $J_i$ in Eq.~\eqref{eq:IsingH1} are the site-specific nearest neighbor coupling of the magnetic moments $s_i$.  This could describe an inhomogeneous magnet with short range interactions.   On the other hand, the parameters $J_\alpha$ in Eq.~\eqref{eq:IsingH2} are the nearest and second nearest neighbor coupling of the magnetic moments describing a homogeneous magnet with long(er) range interactions.  A visualization of the model manifold for Eq.~\eqref{eq:IsingH1} is given in Figure~\ref{fig:Ising_3_2_r_1} for the case of three spins and two parameters.  The model manifold for Eq.~\eqref{eq:IsingH2} is given in Figure~\ref{fig:Ising_4_2_k_1} for the case of two parameters and four spins with periodic boundary conditions.  Because these models are derived from different abstract models, the topological features are different (a square vs.~a triangle).

\begin{figure}
\includegraphics[width=\linewidth]{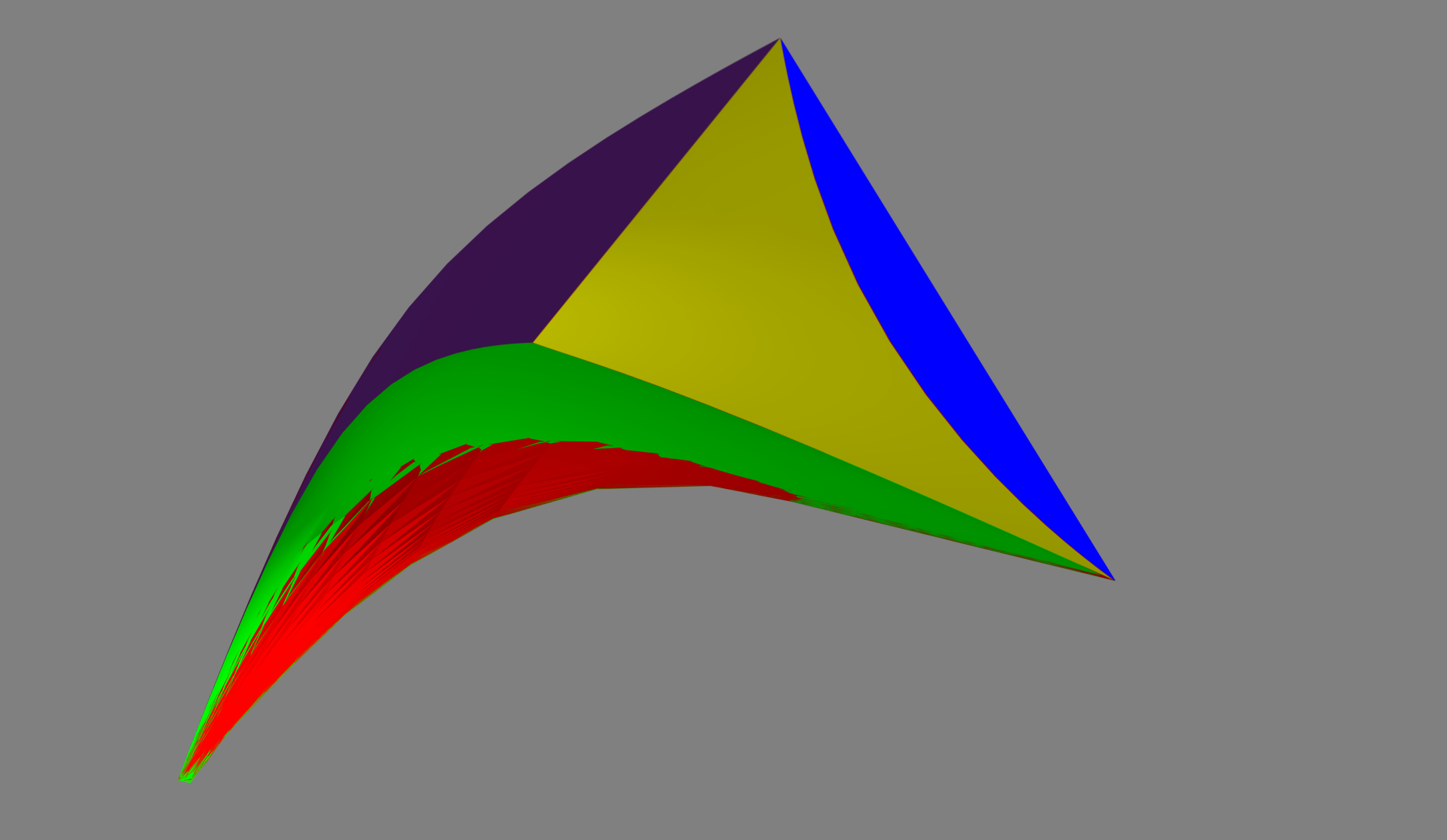}
\caption[]{\label{fig:ESR_1}\textbf{Model Manifold of the Enzyme-Substrate Model} is a three dimensional volume with five sides (colored red, green, blue, yellow, and purple).}
\end{figure}

A final example is drawn from biochemistry, an enzyme substrate reaction: $ \textrm{E} + \textrm{S} \leftrightharpoons \textrm{ES} \rightarrow \textrm{E} + \textrm{P}$.  Modeled as three mass-action reactions, the time-dependence of the concentration of each chemical species is determined by the set of differential equations
\begin{eqnarray}
  \label{eq:ESR}
  \ddt{\E} & = & -k_f \E \Ss + k_r \ES + k_\mathrm{cat} \ES \\
  \ddt{\Ss} & = & -k_f \E \Ss + k_r \ES \\
  \ddt{\ES} & = & k_f \E \Ss - k_r \ES - k_\mathrm{cat} \ES \\
  \ddt{\Prod} & = & k_\mathrm{cat} \ES.
\end{eqnarray}
We take the observations to be the values of each of the chemical species at various times with added random Gaussian noise of variable size.  This model has three parameters, the three reaction rates $k_f$, $k_r$, $k_\mathrm{cat}$, so that the model manifold is a volume rather than a surface.  The model manifold is shown in Figure~\ref{fig:ESR_1} where the five colors are five sides (which we consider below) that enclose the volume.  Initial conditions are chosen to be a known mixture of $\E$, $\Ss$ and $\ES$.  This is an unrealistic initial condition since $\ES$ is unstable and spontaneously decays into its constituent parts.  We consider a more realistic scenario in a later section when we discuss manifold collapse.

To this point, the observations for all of the statistical models were chosen so that all the parameters of the model were identifiable (both structurally and practically).  As such, repeated sampling from these models contains all the information to uniquely identify the parameters in the abstract model, which in turn leads to precise predictions for other observations (i.e., extrapolate to other statistical models).

We now consider for which other observations the structural identifiability of the parameters in the abstract model will be preserved (at least locally).  To answer this question, we use the fact mentioned above above that a statistical model is locally structurally identifiable, if and only if the FIM is non-singular.  Therefore, structural preserving experimental conditions can distort the distances measured on the model manifold provided there are no new singularities introduced into the metric tensor.  These types of transformations are known as diffeomorphisms, i.e., differentiable transformations of the model manifold.  Therefore the manifolds in Figures~\ref{fig:Exps_1}-~\ref{fig:Ising_3_2_r_1} are diffeomorphic to a square, Figure~\ref{fig:Ising_4_2_k_1} is diffeomorphic to a triangle and Figure~\ref{fig:ESR_1} is diffeomorphic to a pentahedron.  

As an aside, throughout this paper we say that a manifold that is square-like (as in Figures~\ref{fig:Exps_1}-~\ref{fig:Ising_3_2_r_1}) is topologically a square (and similarly for other shapes).  By this we mean the manifold is diffeomorphic to a square.  We warn the reader that the phrase ``topological equivalence'' is colloquially used to mean invariance under homeomorphisms rather than diffeomorphisms.  In this paper we restrict ourselves to the study of differential topological properties of manifolds, so we use this phrase without ambiguity although it is not standard.

Notice that each of the manifolds in Figures~\ref{fig:Exps_1}-\ref{fig:ESR_1} are bounded by edges which are in turn bounded by corners.  If we restrict our attention to observations that are diffeomorphic to those in these figures, then this hierarchical structure of boundaries and edges will be preserved.  This indicates that the hierarchical boundary structure is a feature of the abstract model and not of the specific observations.  Indeed, we now give explicit formulas and interpretations for each of these limits and find that these boundaries always represent an extreme limit of the principles in the abstract model.

The simplest example is in Figures~\ref{fig:Exps_1}-\ref{fig:Exps_2} in which the model manifold is topologically a square.  The limits of the four edges correspond to either of the parameters reaching their physically limiting values individually: $\theta_\mu \rightarrow \infty$ or $\theta_\mu \rightarrow 0$.  Because these limits act on the underlying physical principles, it is not difficult to ascribe a physical interpretation to these edges.  They are the cases in which one of the radioactive species either decays instantly or not at all (relative to the experimental time scales).

The generalized Ising model in Eq.~\eqref{eq:IsingH1} (depicted graphically in Figure~\ref{fig:Ising_3_2_r_1}) is likewise diffeomorphic to a square.  The four limiting cases are likewise each of the two parameters reaching the limits of their physically relevant values: $J_\mu \rightarrow \pm \infty$.  These limits physically correspond to perfect, local ferromagnetism or anti-ferromagnetism between two neighboring spins.  

Now consider the generalized Ising model in Eq.~\eqref{eq:IsingH2} and visualized in Figure~\ref{fig:Ising_4_2_k_1}.  This model manifold is diffeomorphic to a triangle.  Because this model is constructed to be translationally invariant, it is natural to consider a Fourier transform of the spins as in\cite[supplement]{transtrum2014model}.  One limiting case corresponds to the limit of $J_1 \rightarrow \infty$, $J_2 \rightarrow - \infty$ such that $J_1 + 2 J_2 \rightarrow \textrm{finite}$.  Careful analysis shows that in this limit, configurations with the highest Fourier frequency have infinite energy (i.e., have zero probability).  We refer to this as the ferromagnetic limit (high-frequency configurations in the spins correspond to anti-ferromagnetic order and have zero probability).  Another limit occurs when $J_1 -\rightarrow \infty$, $J_2 \rightarrow - \infty$ such that $J_1 - 2 J_2 \rightarrow \textrm{finite}$ which in a similar way removes the low-frequency configurations in the spins, i.e.,  an anti-ferromagnetic limit.  Finally, the limit $J_2 \rightarrow  \infty$ with $J_1$ remaining finite corresponds to the limit of no mid-frequency configurations.

Finally, we consider the five surfaces bounding the pentahedron in Figure~\ref{fig:ESR_1}.  The green surface corresponds to the limit $k_r \rightarrow 0$, i.e., the first reaction is no longer reversible.  The red surface corresponds to $k_f, k_r \rightarrow \infty$ such that $K_d = k_r/k_f$ remains finite, interpreted as the case in which the first reaction is always in equilibrium.   The blue surface corresponds to $k_\mathrm{cat} \rightarrow 0$, i.e., the second reaction does not occur.  The yellow surface corresponds to $k_f \rightarrow 0$, i.e., the first reaction proceeds only in the reverse direction.  The purple surface correspond to the limit $k_r, k_\mathrm{cat} \rightarrow \infty$ such that the ratio $k_r/k_\mathrm{cat}$ remains finite, corresponding to the case that the intermediate complex $\ES$ never accumulates.   The topological relationship among the faces, edges, corners, etc. of the model is rich with physical meaning as we will see in section \ref{sec:ModelClasses}.

\section{Manifold Collapse leads to changes in model structure}
\label{sec:collapse}

We now consider how the analysis described in section~\ref{sec:InformationTopology} changes when the observations alter the topological structures of the manifolds, i.e., are not related by diffeomorphisms.  Since the models in section~\ref{sec:InformationTopology} were constructed to be identifiable, a structural change here corresponds to a \emph{coarsening} of the observations.  In most cases, this means simply not observing some of the potential model predictions.  In the language of mathematical probability, this means the distribution is marginalized over a subset of the random variables resulting is a coarse-grained set of observations.  Geometrically, the corresponding manifolds are compressed along some direction.  In this way, manifolds may be compressed, folded, or edges may be glued together.  On the other hand, reversing this process can result in manifolds that are stretched or torn.

Consider the exponential model in Eq.~\eqref{eq:ExpDef} for the limit in which the difference in the exponential terms are not observed.  This occurs if the specific radioactive products of the two decay channels are indistinguishable, or if the experimental cost of observing the difference is prohibitive so that the experiment is not performed.  In this case, Eq.~\eqref{eq:ExpDef} is modified to be
\begin{equation}
  \label{eq:ExpDef_CG}
  \xi_i(\theta_1, \theta_2)  =  
      e^{-\theta_1 t_i} + e^{-\theta_2 t_i} + z_i.
\end{equation}

\begin{figure}
\includegraphics[width=\linewidth]{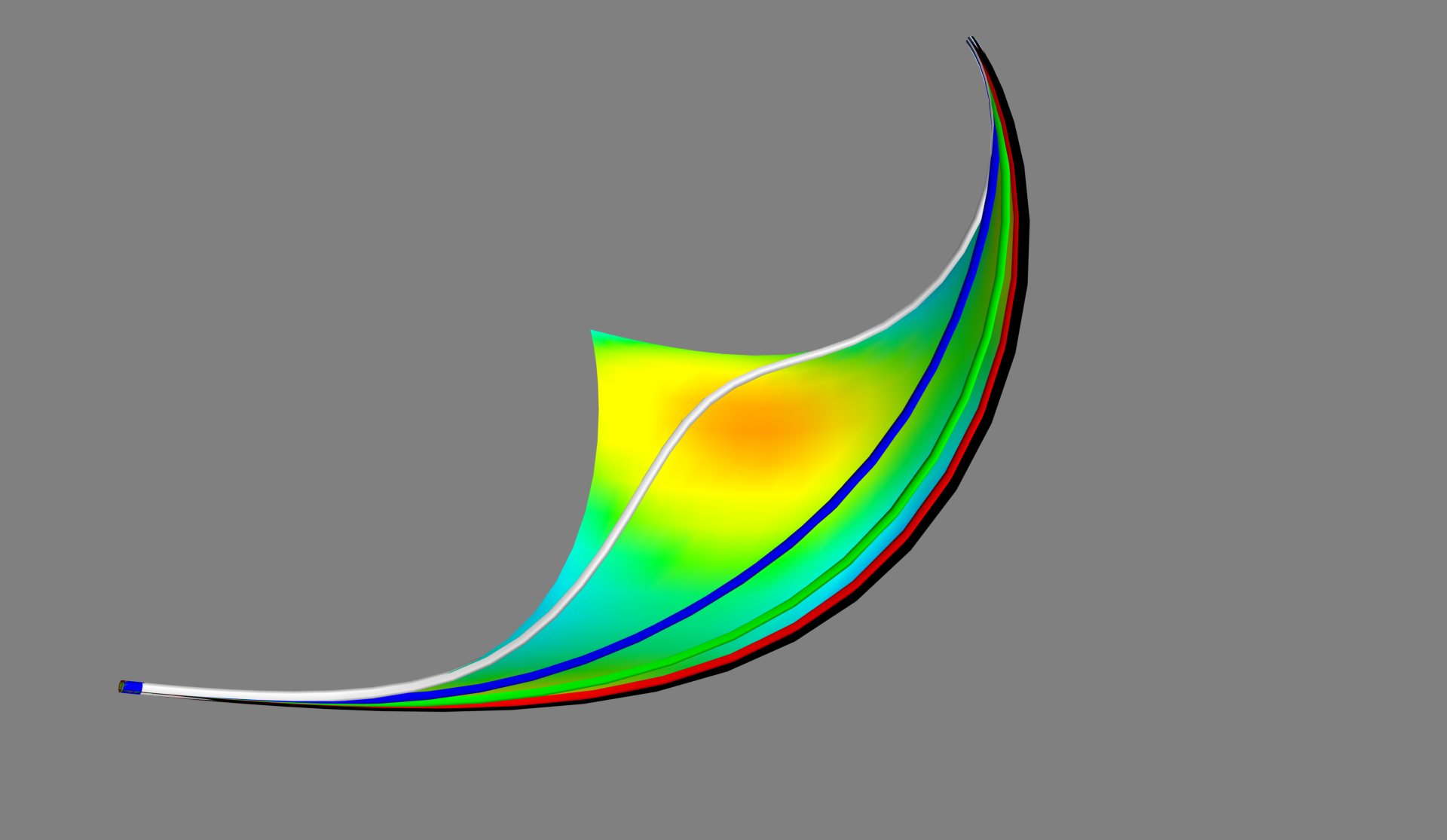}
\caption[]{\label{fig:Exps_CG_2} \textbf{Coarse-grained Exponential Model}.  If only the sum of the exponential terms is observed, then the model manifold is structurally different from that in Figures~\ref{fig:Exps_1} and \ref{fig:Exps_2}; it is now a triangle.  The topological change reflects a change in the information content of the observations about the underlying theory and manifests itself as a structural unidentifiability in the resulting model.}
\end{figure}

\begin{figure}
\includegraphics[width=\linewidth]{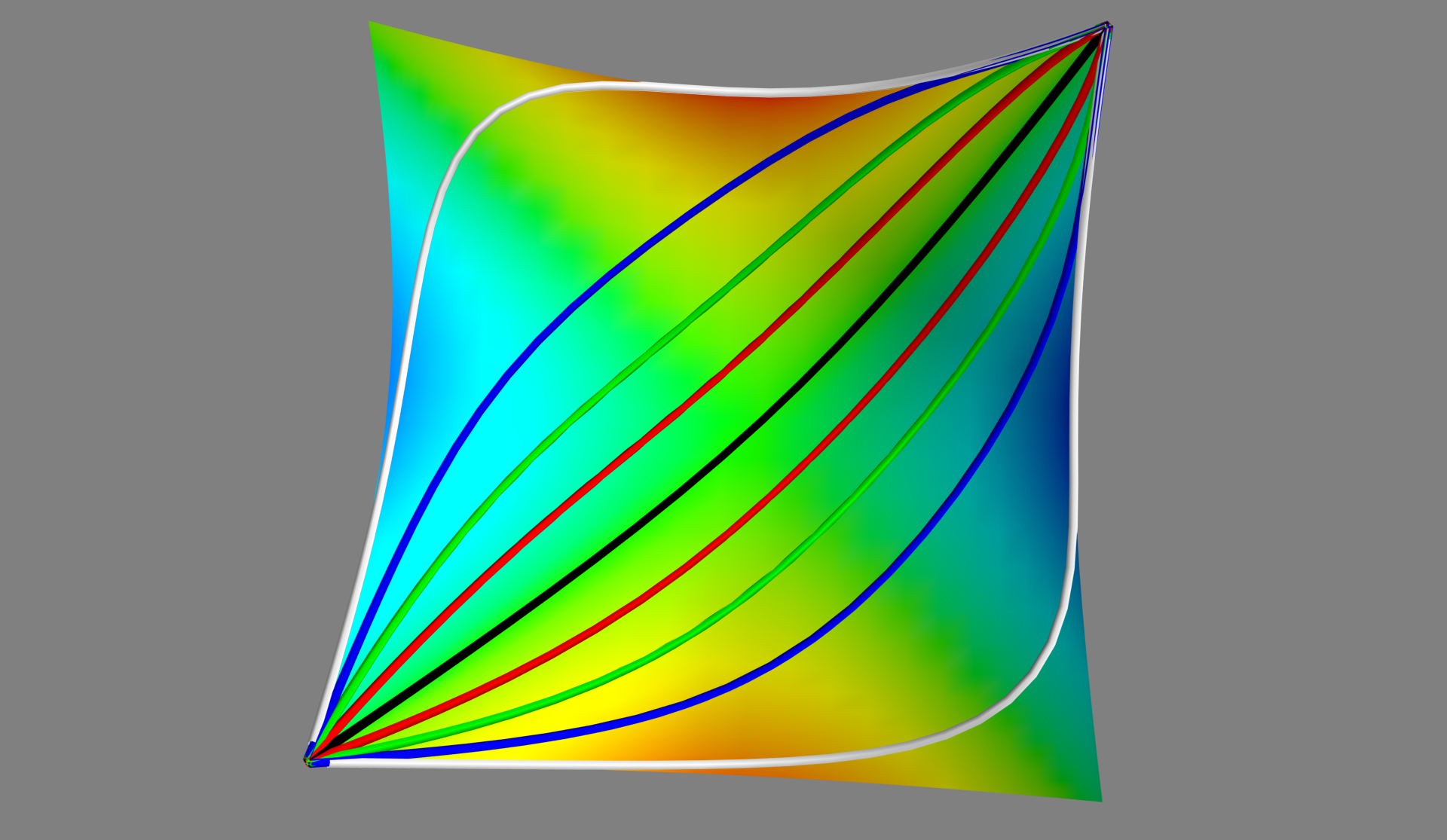}
\caption[]{\label{fig:Exps_CG_1}\textbf{Full Exponential Model} showing which points will be identified under coarse-graining.  In transitioning from the manifolds in Figures~\ref{fig:Exps_1} and \ref{fig:Exps_2}, the (square) manifold is effectively folded in half to produce the triangle structure in Figure~\ref{fig:Exps_CG_2}.  Here, lines of the same color on the original topology become identified with each other after coarsening.  The black line corresponds to the fold line.}
\end{figure}

The manifold of this model is given in Figure~\ref{fig:Exps_CG_2}.  Notice that the topological structure of the manifold is fundamentally different from that in Figures~\ref{fig:Exps_1} and \ref{fig:Exps_2}.  The relationship between the coarsened manifold and the original manifold is illustrated by the colored lines in Figures~\ref{fig:Exps_CG_2} and \ref{fig:Exps_CG_1}.  In effect, the manifold has been folded in half such that the two white lines in Figure~\ref{fig:Exps_CG_1} are identified with each other and correspond to the white line in Figure~\ref{fig:Exps_CG_2}.  Similarly for the blue, green and red lines.  The black line is the ``fold line.''

The black ``fold line'' in Figures~\ref{fig:Exps_CG_2} and \ref{fig:Exps_CG_1} corresponds to the curve for which $\theta_1 = \theta_2$ in the model.  This line is significant because for the coarsened model in Eq.~\eqref{eq:ExpDef_CG}, the Fisher Information is singular along this line.  This singularity is the mathematical indication of the corresponding structural change.  For all other points on the model, the Fisher Information remains nonsingular, so that the model is still structurally identifiable in the local sense almost everywhere.  However, the model is no longer globally identifiably since each point on the manifold corresponds to two points in parameter space.  

The structural change illustrated in Figures~\ref{fig:Exps_CG_2} and \ref{fig:Exps_CG_1} indicates that the observations carry qualitatively different information about the parameters of the abstract model.  Specifically, there is a loss of information regarding the distinguishability of the radioactive products.  In order to construct an identifiable model, the abstract model must be modified.  In this case, the physical domain of the parameters can be restricted to $\theta_1 > \theta_2$, i.e., we arbitrarily order the parameters so that they are no longer identified with a specific radioactive agent.

\begin{figure}
\includegraphics[width=\linewidth]{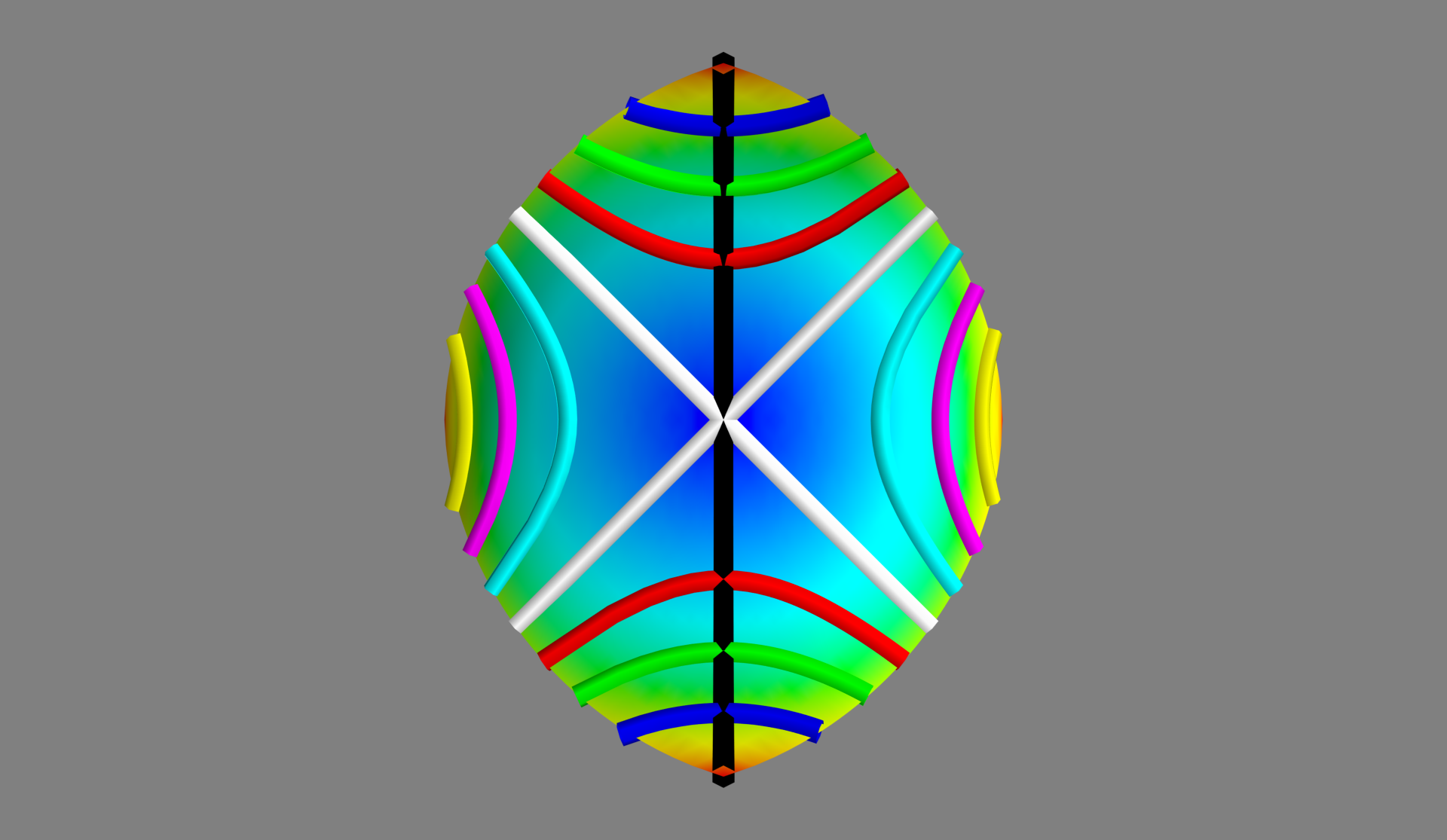}
\caption[]{\label{fig:Ising_3_2_r_CG}\textbf{Identified points for the generalized Ising model in Eq.~\eqref{eq:IsingH1}.}  After coarsening, the two dimensional square topology collapses to a line segment.  Here, lines of the same color are the set of points that collapse to an single point on the final line segment.  Notice that this collapse involves both a type of compression, in which a sequence of connected points are squeezed together, and a global folding, in which blue lines (for example) on opposite sides of the manifold are identified with each other.  The black line represents the model in which the two parameters are equal, i.e., the usual one-parameter nearest neighbor Ising model.  Notice that this line is folded in half under coarsening.}
\end{figure}

Now consider the generalized Ising model in Eq.~\eqref{eq:IsingH1}.  For the specific case of three spins visualized in Figure~\ref{fig:Ising_3_2_r_1}, we now consider the effect of observing only spins one and three and marginalizing the distribution over spin two.  In this case, the two dimensional manifold in Figure~\ref{fig:Ising_3_2_r_1} collapses to a one-dimensional curve.  The nature of this collapse is illustrated by the colored lines in Figure~\ref{fig:Ising_3_2_r_CG}.  After marginalizing the distribution, the colored lines each collapse to a single point.  Additionally, the manifold is ``folded up'' so that the disconnected lines of the same color each map to the same point.  That is, the blue line near the top and near the bottom each collapse to a single point and the manifold is folded in half to identify these points.

It is interesting to consider the effect on the common Ising model corresponding to $J_1 = J_2$ in Eq.~\eqref{eq:IsingH1}.  This one-dimensional curve is illustrated by the black line in Figure~\ref{fig:Ising_3_2_r_CG}.  Notice that it is folded in half by the coarse-graining.  

Since the previously two-dimensional manifold is collapsed to a one-dimensional curve, the FIM matrix is singular for all parameter values upon coarsening.  Thus the new model is structurally unidentifiable.  The meaning of the parameters in the abstract model must be modified in order to construct an identifiable model.  In this case, the information lost is the nature of the short-range interactions.  Are they ferro-magnetic or anti-ferromagnetic?  After coarsening, the answer to this question is lost.  As shown in reference\cite{transtrum2014model}, the new model can be exactly represented by an effective interaction between spins one and three:
\begin{equation}
  \label{eq:IsingH1_CG}
  H = -\tilde{J} s_1 s_3.
\end{equation}
We quantify this effective interaction with the renormalized parameter $\tilde{J}$.  The origin of this interaction is understood to be mediated by the microscopic interactions described by $J_1$ and $J_2$ neither of which can be identified individually from the data.

\begin{figure}
\includegraphics[width=\linewidth]{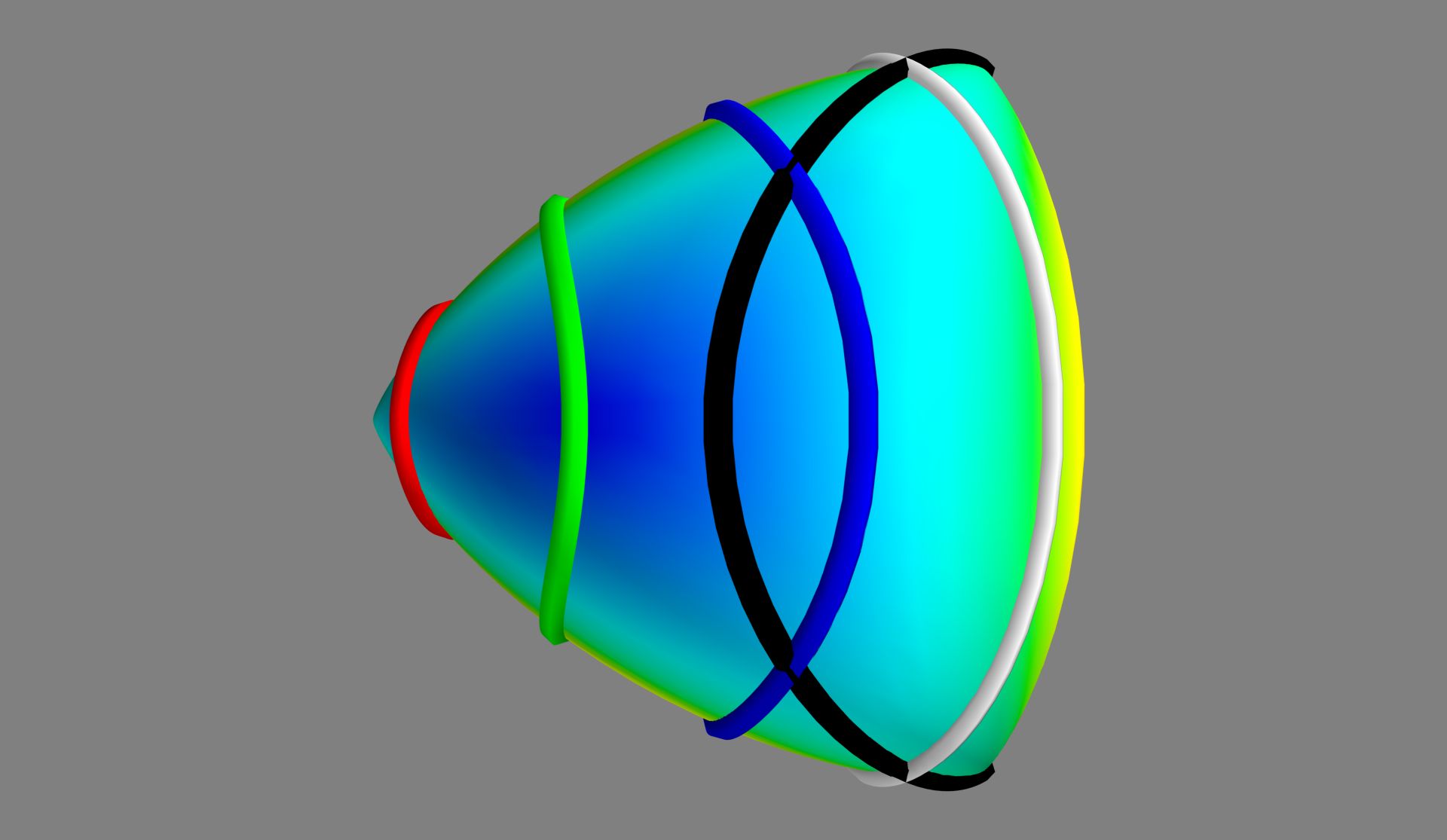}
\caption[]{\label{fig:Ising_4_2_k_CG}\textbf{Identified points for generalized Ising model in Eq.~\eqref{eq:IsingH2}.}  After coarsening, the two dimensional triangle topology condenses to the same line segment as that for the model in Eq.~\eqref{eq:IsingH1}.  Here, lines of the same color are collapsed to a single point.  Unlike the collapse in Figure~\ref{fig:Ising_3_2_r_CG}, the structure does not experience any folding, rather the area between two of the edges are collapsed to a line and the third edge collapses to a point.  The black line corresponds to the model with $J_2 = 0$, with $J_1$, i.e., the usual one-parameter nearest neighbor Ising model.  This black line is the same as that in Figure~\ref{fig:Ising_3_2_r_CG}.  Although the whole manifold is not folded, this particular sub-manifold is folded in half under coarsening, just as it was in Figure~\ref{fig:Ising_3_2_r_CG}.}
\end{figure}

The generalized Ising model in Eq.~\eqref{eq:IsingH2} also collapses if we similarly only observe spins one and three, ignoring spins two and four.  The two dimensional surface in Figure~\ref{fig:Ising_4_2_k_1} then collapses to a one dimensional curve.  The details of this collapse are illustrated in Figure~\ref{fig:Ising_4_2_k_CG} in which lines of a single color collapse to a single point.  Notice that the manifold is not ``folded'' in the same way as that in Figure~\ref{fig:Ising_3_2_r_CG}.  Indeed, two edges of the triangle are simply brought together and the third edge is collapsed to a point.  Considering the curve $J_2 = 0$ (given by the black line), we see that this curve is folded in half, just as it was in Figure~\ref{fig:Ising_3_2_r_CG}.

The two abstract models of these two generalized Ising models are fundamentally different.  Eq.~\eqref{eq:IsingH1} corresponds to a inhomogeneous magnet with short-range interactions, while Eq.~\eqref{eq:IsingH2} is a homogeneous magnet with longer-range interactions.  These differences are captured in the different topological structures.  However, upon coarsening, both models collapse to a one-dimensional curve.  In fact, they collapse to the \emph{same} model and are realizations of the same abstract model.  This effective model is one in which observed spins interact ``directly''.  That ``direct'' interaction, is of course, mediated by different microscopic interaction, of a now unknown nature (and unknowable from the coarsened observations).

This simple example demonstrates how manifold collapse can both explain universality and justify the use of simple, effective models.  This example is best understood in analogy with similar arguments based on renormalization group analysis.  The similarity of these analyses becomes a vehicle for generalizing these concepts to broader model classes.

\begin{figure}
\includegraphics[width=\linewidth]{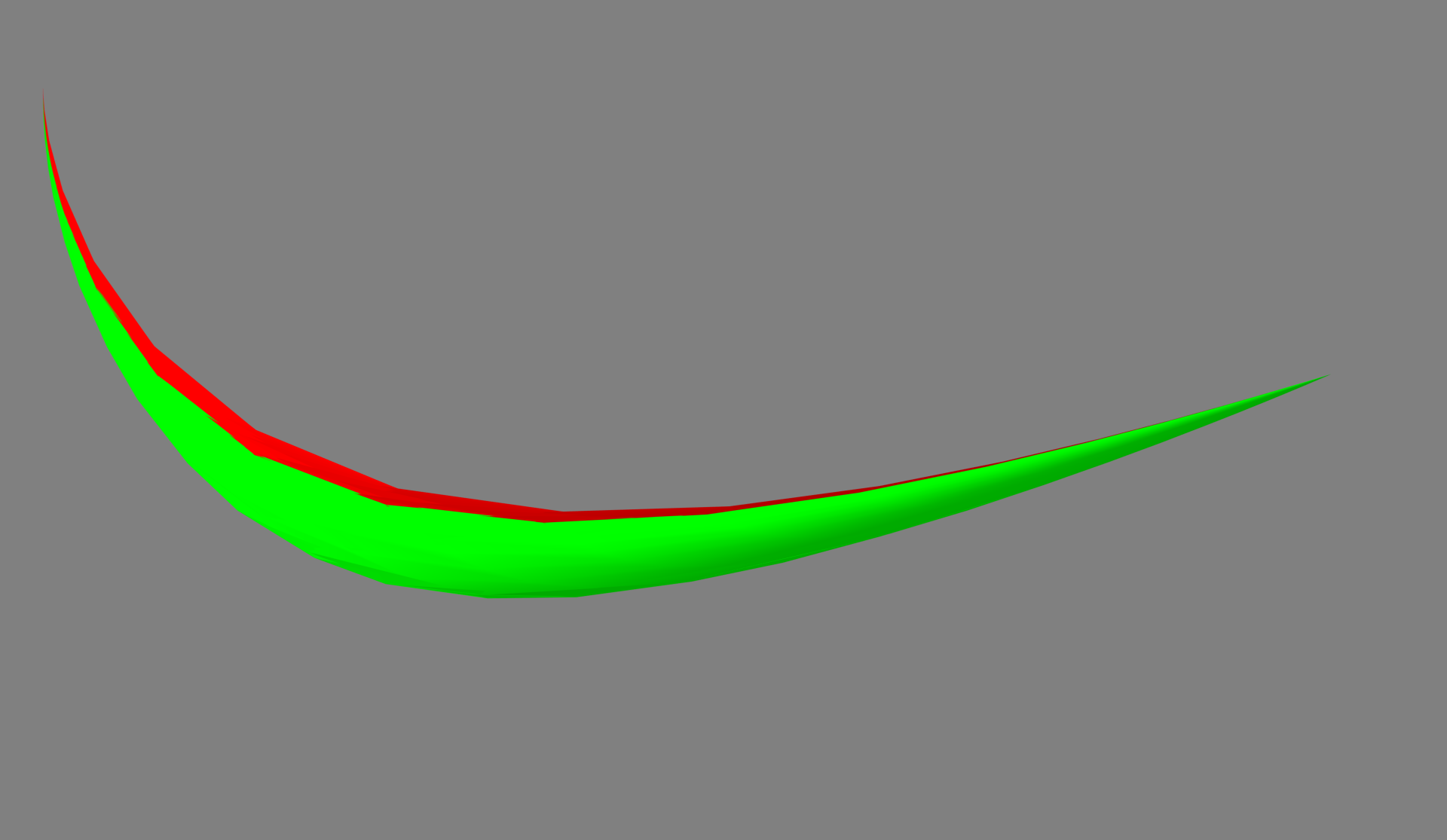}
\caption[]{\label{fig:ESR_CG} \textbf{Coarse-grained Enzyme-Substrate Model}  A more realistic experimental conditions for the enzyme-substrate model results in a manifold that has not completely collapsed as it remains a three dimensional volume.  However, several of the boundaries of the manifold in Figure~\ref{fig:ESR_1} has collapsed.  The yellow and blue surfaces have collapsed to points and the purple surface has collapsed to a line.  The resulting model is structurally identifiable, but not practically identifiable.  The practical unidentifiability is closely related to the topological changes of the boundary structure.}
\end{figure}

Finally, we turn our attention the enzyme-substrate model.  To coarsen this model, we consider an alternative initial condition in which only $\E$ and $\Ss$ have nonzero initial condition and observe only the final product $\Prod$ (ignoring the time course of the other three variables).  The corresponding model manifold is illustrated in Figure~\ref{fig:ESR_CG}.  The model manifold for these observations is three dimensional; however, the pentahedron of Figure~\ref{fig:ESR_1} has collapsed to a narrow volume bounded by two surfaces (red and green), each of which are digons, i.e., two-sided polygons.

To understand manifold collapse in Figure~\ref{fig:ESR_CG}, Notice that if $k_\mathrm{cat} = 0$ in Eqs.~\eqref{eq:ESR}, that no product can be produced.  Consequently, the blue and yellow surfaces in Figure~\ref{fig:ESR_1} are collapsed to a single point in Figure~\ref{fig:ESR_CG}.  The entire manifold has been compressed, so that the two remaining surfaces are very close to one another and are therefore \emph{practically} unidentifiable.  Thus, in principle it is possible to identify all of the parameters in the model using only observations of the product; however, in practice the near-collapse of the manifold makes it very difficult.

The structural collapse of the boundaries in Figure~\ref{fig:ESR_CG} reflects the causal dependence among $k_f$, $k_r$, and $k_\mathrm{cat}$ in the abstract model.  The practical unidentifiability resulting from the coarsening is intimately tied to this relationship and is manifest as a structural change in the model's topology.  Inspecting Figure~\ref{fig:ESR_1}, we see that the blue and yellow surfaces are necessary to ``pull apart'' the red and green surfaces.  If the choice of observations collapses the blue and yellow surface to a single point (as in Figure~\ref{fig:ESR_CG}), it becomes difficult to statistically distinguish between the red and green models (as well as all the models between them).  We discuss this phenomenon in more detail in section~\ref{sec:ModelClasses}.  

\section{Hyper-corners Define phenomenological classes}
\label{sec:ModelClasses}

\subsection{Boundary structures are represented by Hasse Diagrams}
\label{sec:Hasse}

\begin{figure}
\includegraphics[width=\linewidth]{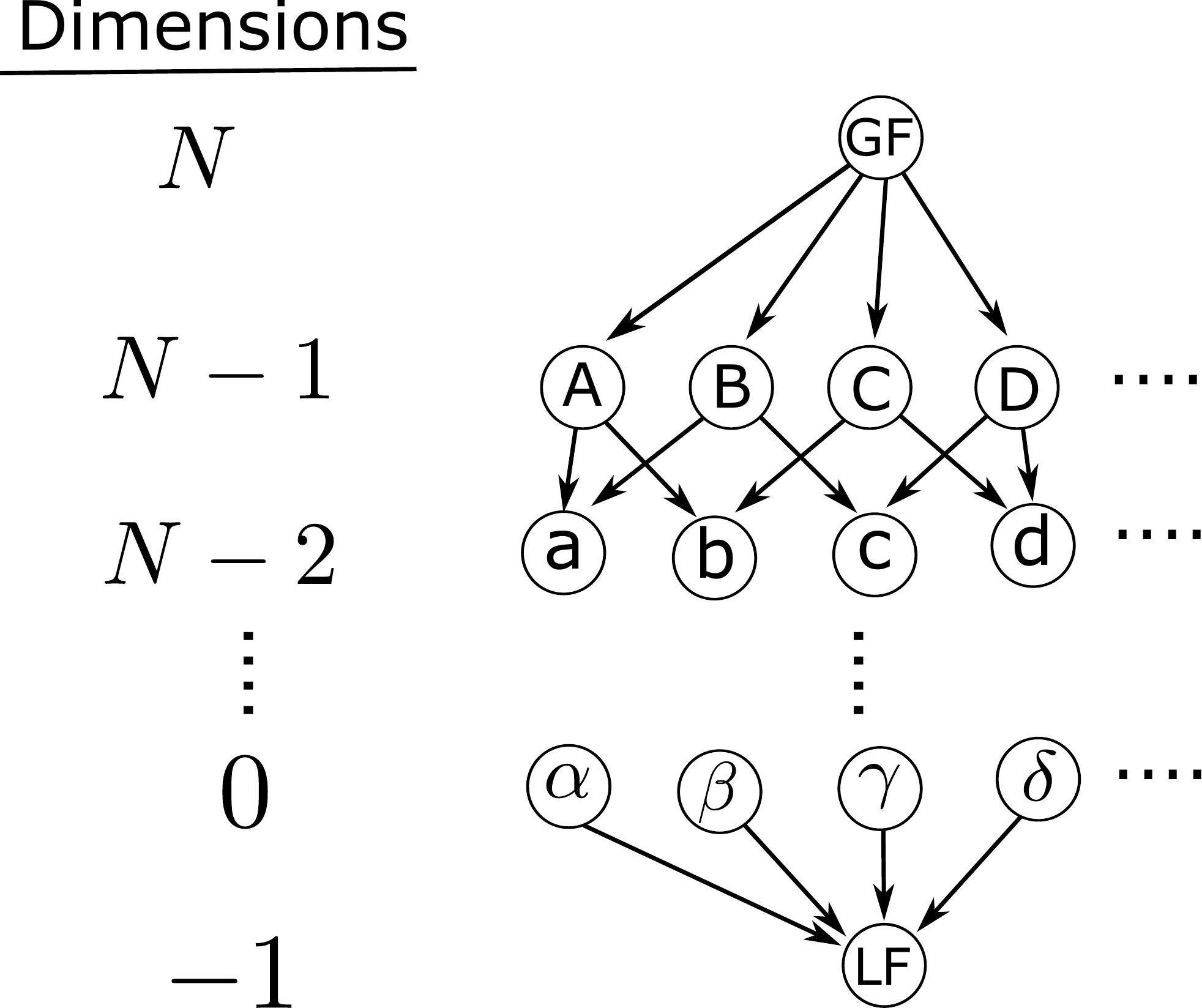}
\caption[]{\label{fig:TreeCartoon} \textbf{Hasse Diagram.}  The relationship among boundaries, corners, hyper-corners, etc.~may be summarized by a hierarchical graph structure known as a Hasse Diagram.  The top node of the graph represents the full $N$-dimensional model (known as the greatest face and denoted by GF).  The first row of nodes correspond to the surfaces of dimension $N-1$ that bound this model.  The next row represents the $N-2$-dimensional surfaces that bound those in the previous row, and so forth.  The tips of the graph correspond to models of dimension zero (i.e. a single point).  It is common in Hasse diagrams to have a single node at the bottom (corresponding to dimension -1) representing the empty set and known as the least face (denoted by LF).}
\end{figure}

The models we have considered here have manifolds with an intrinsic hierarchical structure.  They are bounded by (hyper-)surfaces of one less dimension, which are in turn bounded by other (hyper-)surfaces of even lower dimension.  This structure is derived from the ``rules'' of the abstract model.  It has been shown that this structure is common to many models\cite{transtrum2016manifold}.  This hierarchical structure is described mathematically as a graded partially ordered set, i.e., a graded poset.  The grading of each element in the set is determined by its dimension (corners are zero-dimensional, edges are one-dimensional, etc.).  The partial ordering is induced by the fact that some corners are contained within some edges but not others.  In general models that share this basic construction will have many more than two or three parameters making them difficult to visualize in data space as we have done previously.  However, the relationships among this hierarchy of hyper-surfaces can be represented as a hierarchical graph structure known as a Hasse diagram, as shown in Figure~\ref{fig:TreeCartoon}.  

A Hasse diagram graphically illustrates the relationship among the model manifold, its faces, edges, corners, etc.  The complete model is represented as a single manifold of dimension $N$ (where $N$ is the number of parameters), given by the top node of the graph.  This manifold may be bounded by a collection of hyper-surfaces of dimension $N-1$, i.e., the second row in the figure.  Likewise, each of these hyper-surfaces may be bounded by other hyper-surfaces of dimension $N-2$, i.e., the third row in the figure.  In this paper, we do not consider models that are unbounded in some directions.

The arrows connecting the top node to the nodes in the first row show that each of these surfaces is a boundary to the node above it.  The arrows connecting the second and third rows similarly represent which $N-2$ dimensional hyper-surfaces are boundaries to which $N-1$ dimensional hyper-surfaces.  In this way, a Hasse diagram summarizes the topological relationships among all the boundaries of the model.  

\begin{figure}
\includegraphics[width=\linewidth]{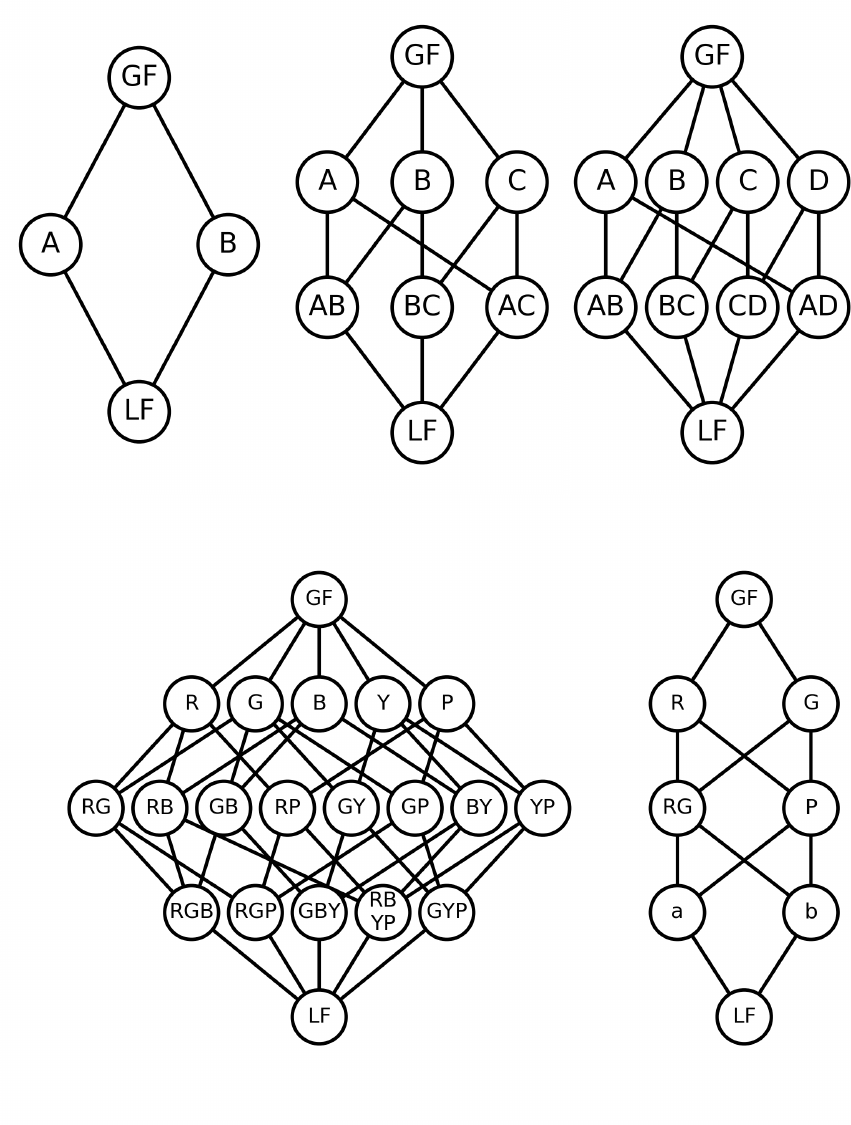}
\caption[]{\label{fig:hassediagrams} \textbf{Examples of Hasse Diagrams.}  Top: From left to right the Hasse diagrams of a line segment, a triangle, and a square.  In each case, edges are arbitrarily labeled by a letter and corners by two letters corresponding to the adjacent edges.  Bottom left: The Hasse diagram of the pentahedron in Figure~\ref{fig:ESR_1}.  Faces are labeled with the letter of the corresponding color in Figure~\ref{fig:ESR_1} (R for red, etc.).  Four of the five faces are triangles; the green face is a square.  Bottom right: The Hasse diagram for Figure~\ref{fig:ESR_CG}.  In this case, the three dimensional volume is bounded by two digons.}
\end{figure}

Near the bottom of the graph are nodes of zero dimension (labeled by Greek letters) which are vertices of the manifold.  It is common in the theory of abstract polytopes to include a single node in the Hasse diagram below the points of dimension 0 (i.e., dimension -1) corresponding to the empty set\cite{grunbaum1967convex,ziegler1995lectures,brondsted2012introduction}.  Examples of Hasse diagrams for the manifolds in sections~\ref{sec:InformationTopology}-~\ref{sec:collapse} are shown in Figure~\ref{fig:hassediagrams}.

\subsection{$f$-vector and Euler Characteristic}
\label{sec:euler}

A topologically important quantity that can be read directly from the Hasse diagram is the $f$-vector.  The $f$-vector is a list of integers giving the total number of nodes on each row (i.e., of a particular dimension) of the Hasse diagram.  The $f$-vectors for the Hasse diagrams in Figure~\ref{fig:hassediagrams} are line segment: (1,2,1), triangle: (1,3,3,1), square: (1,4,4,1), enzyme reaction: (1,5,8,5,1), coarsened enzyme reaction: (1,2,2,2,1).  

The Euler-characteristic is a topological invariant calculated as the alternating sum of terms in the $f$-vector
\begin{equation}
  \label{eq:EulerCharacteristic}
  \chi = \sum_{i=0}^{N-1} (-1)^i f_i.
\end{equation}

\begin{figure}
\includegraphics[width=\linewidth]{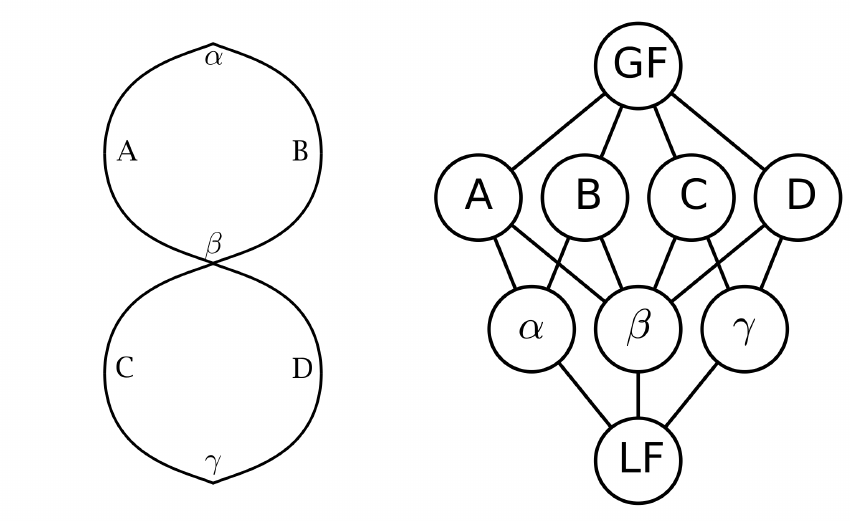}
\caption[]{\label{fig:2Digons} \textbf{Collapsed Exponential Model.} Left: If only the difference in exponentials is observed (as in Eq.~\eqref{eq:ExpDef_CG2}), then the square topology in Figure~\ref{fig:Exps_1} collapses to a point along the line $\theta_1 = \theta_2$. The manifold structure breaks down at the vertex $\gamma$, so this structure is not a manifold.  It is two separate manifolds, each with the topology of a digon connected at their tips.  Right: the Hasse diagram for this structure.}
\end{figure}

It is straightforward to check that for each of the models considered here, $\chi = 1 - (-1)^N$\cite{grunbaum1967convex}.  The significance of this result is that all of the manifolds we consider here are orientable and none have holes or handles (such as would a donut or a coffee mug).  We anticipate that these are properties that will be common for many models, but it is possible to find other examples.  For example, consider another coarse-grained version of Eq.~\eqref{eq:ExpDef} in which only the difference of exponentials is observed:
\begin{equation}
  \label{eq:ExpDef_CG2}
  \xi_i(\theta_1, \theta_2)  =  
      e^{-\theta_1 t_i} - e^{-\theta_2 t_i} + z_i.
\end{equation}
In this case, the square of Figure~\ref{fig:Exps_1} is pinched off to a point at the line $\theta_1 = \theta_2$.  The resulting structure is not a manifold since the manifold structure breaks down at $\theta_1 = \theta_2$.  Instead, it is two manifolds that are topologically digons that are glued together at one of their tips as illustrated in Figure~\ref{fig:2Digons}.  The $f$ vector for this structure is (1,3,4,1) and has an Euler characteristic $\chi = -1$.

Because there is only one greatest face and least face, the first and last entries of the $f$-vector will generically be one, there are several equivalent variations of Eq.~\eqref{eq:EulerCharacteristic}.  For models satisfying $\chi = 1 - (-1)^N$, we can similarly write\cite{grunbaum1967convex}
\begin{eqnarray}
  \label{eq:euler2}
  \sum_{i=0}^N (-1)^i f_i & = & 1 \\
  \label{eq:euler3}
  \sum_{i=-1}^N (-1)^i f_i & = & 0.
\end{eqnarray}

\subsection{ Hasse diagrams identify important parameter combinations}
\label{sec:hierarchyparameters}

Another use of the Hasse diagram is understanding the relationship between model parameters and the model's phenomenology.  Quite often, particularly for sloppy or practically unidentifiable models, the $N$ independent parameter combinations of a model are not equally important for explaining the model's observations, nor is it clear how variations in parameter values translate to model behavior.  Indeed, for many observations, there is a clear hierarchy of importance in the model parameters that is revealed by an eigenvalue decomposition of the FIM.  The eigenvectors of the FIM can then be interpreted as the linear combinations of parameters that are relatively important or unimportant for understanding the model behavior.  Unfortunately, this interpretation is based on a local, linear analysis.  In reality, the truly important parameters combinations are nonlinear combinations of the model's bare parameters.  Identifying and interpreting the appropriate nonlinear combination requires a global, topological analysis.  To illustrate, we consider the enzyme-substrate model in Figure~\ref{fig:ESR_CG}.


Notice that the coarsened manifold in Figure~\ref{fig:ESR_CG} is a three dimensional volume.  As such, the statistical model is structurally identifiable.  However, the manifold is very thin and as such it is practically unidentifiable.  The practical unidentifiability is closely related to the collapse of the boundaries.  Furthermore, there is a global anisotropy; manifold has a clear long axis.

By considering the statistical model for observations of the product P, any information about the rates $k_f$ and $k_r$ of necessity must be inferred through the third parameter $k_\mathrm{cat}$.  Furthermore, when only the substrate is stimulated, any information about this stimulation must first pass through the intermediate complex $\ES$.  It is precisely this informational dependence that is described by the boundary collapse as discussed in section~\ref{sec:collapse}.  Because the boundaries of the manifold collapse, many different points on the manifold are drawn very near one another leading to a hierarchy of importance in the parameters.  Parameters that were originally easy to distinguished become practically unidentifiable. 

We now consider the functional form of the red and green surfaces in Figure~\ref{fig:ESR_CG}.  The green surface corresponds to the limit $k_r \rightarrow 0$, i.e., the approximation that the first reaction is not reversible.  The red surface corresponds to the limit $k_f, k_r \rightarrow \infty$ with $K_d = k_r/k_f$ remaining finite,i.e., the approximation that the first reaction is in equilibrium.  From the equilibrium approximation, one can derive the well-known Michaelis-Menten approximation:
\begin{equation}
  \label{eq:MichaelisMenten}
  \ddt{\Prod} = \frac{k_\mathrm{cat} E_0 \Ss}{K_d + \Ss},
\end{equation}
where $E_0 = \E + \ES$ is the total amount of enzyme and $K_d = k_r/k_f$.  

The two edges that join the red and green surfaces can then be found as limiting approximations to Eq.~\eqref{eq:MichaelisMenten}.  They are $K_d \rightarrow 0$, which corresponds to the approximation that that the reaction is always saturated, i.e., a constant production rate:
\begin{equation}
  \label{eq:MichaelisMentenSaturated}
  \ddt{\Prod} = k_\mathrm{cat} E_0.
\end{equation}
Alternatively, the second limit corresponds to $k_\mathrm{cat}, K_d \rightarrow \infty$ with $k_\mathrm{eff} = k_\mathrm{cat}/K_d$ remaining finite leading to the form
\begin{equation}
  \label{eq:MichaelisMentenLinear}
  \ddt{\Prod} = k_\mathrm{eff} E_0 \Ss,
\end{equation}
which corresponds to a linear approximation.  

The vertices bounding the two edges are limits $k_\mathrm{cat}, k_\mathrm{eff} \rightarrow 0$ or $k_\mathrm{cat,} k_\mathrm{eff} \rightarrow \infty$ corresponding to either no reaction or a reaction that that completes instantaneously.  

\begin{figure}
\includegraphics[width=\linewidth]{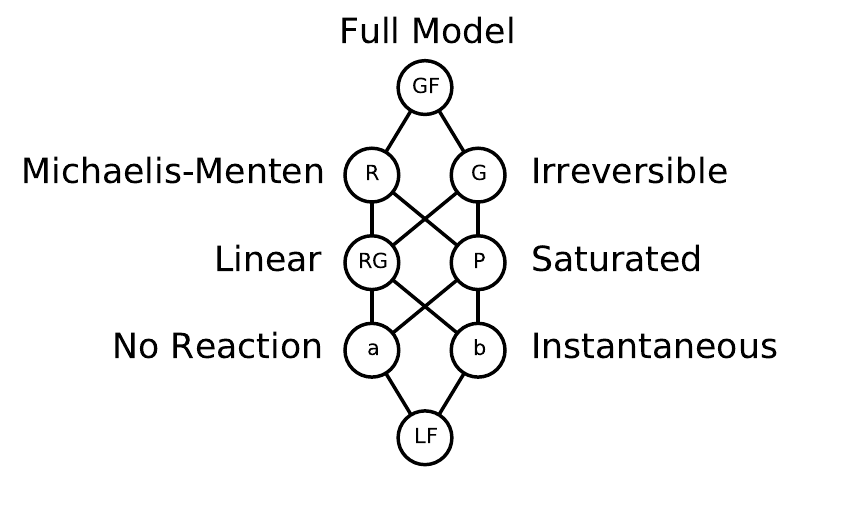}
\caption[]{\label{fig:ESRCG_Hasse} \textbf{Model Hierarchy for coarsened Enzyme-Substrate Model.} The nodes of the Hasse diagram for the coarsened enzyme-substrate model (as in Figure~\ref{fig:ESR_CG}) correspond to a well-known hierarchy of models, including the well-known Michaelis-Menten rate-law.  }
\end{figure}

This hierarchy of models is a standard interpretation of the Michaelis-Menten dynamics in biochemistry and is naturally recovered by interpreting the nodes in the Hasse diagram as illustrated in Figure~\ref{fig:ESRCG_Hasse}.

Notice that a simplified model can be associated with each node in the Hasse diagram.  As one moves to lower dimension, these models become progressively simpler and the relationship between the model behavior and the model parameters becomes progressively clearer.  Furthermore, the simpler models naturally group the parameters of the complex abstract model into the appropriate nonlinear combinations that are connected to the model behavior.  We refer to these simplified models as emergent model classes and their behaviors as the dominant behavioral modes.  The behavior of the full model can then be reinterpreted as a combination of these characteristic modes.

For models of moderate complexity, constructing the Hasse diagram may reveal important insights regarding the informational relationships among model parameters and behaviors such as we have done above for the enzyme-substrate model.  Such insights may guide experimental design, model reduction, and model interpretation.  For more complex models, it may be impractical to explicitly construct the entire diagram.  In these cases, much of the benefit can still be found by identifying a single path down the graph from the top node to an appropriate approximate model.  


\subsection{The observation semi-group and diffeomorphism subgroups}
\label{sec:observationsemigroup}

Observations that preserve the boundary structure of the model manifold are represented by the same Hasse diagram and are related by diffeomorphisms.  The set of diffeomorphisms of a manifold form a group.  Therefore, among the set of all possible observations there are subsets that are groups and characterized by a unique Hasse diagram.  The entire set of all possible experimental conditions do not form a group, however.  This is because some observations lead to manifold collapse which has no unique inverse operation.  All possible observations therefore form a semi-group (a group-like structure but with some elements that do not have an inverse).  This structure is reminiscent of the renormalization group which is also famously not a group, but a semi-group for similar reasons.  We refer to this as the \emph{observation semi-group} and the proper subgroups characterized by unique Hasse diagrams as \emph{diffeomorphism subgroups}.  

The set-like structure of observations induces a partial ordering on the diffeomorphism subgroups.  Let $G_1$ and $G_2$ represent two diffeomorphism subgroups of an abstract model.  We say $G_1 \prec G_2$ if there exist observations $o_1 \in G_1$ and $o_2 \in G_2$ such that $o_1 \subset o_2$.  

\begin{figure}
\includegraphics[width=\linewidth]{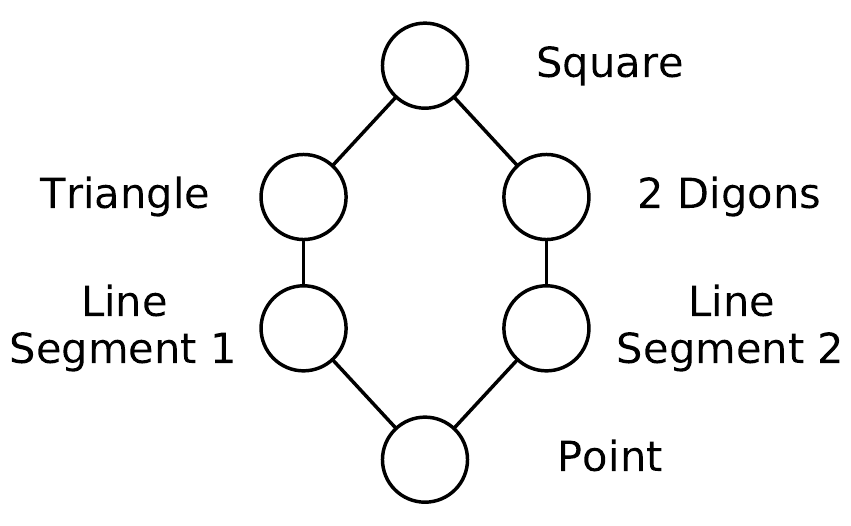}
\caption[]{\label{fig:HasseGroup} \textbf{Hasse Diagram for Observation semi-group}  The model defined in Eq.~\eqref{eq:ExpDef} corresponds to the diffeomorphism subgroup described by a square topology.  The ``triangle'' subgroup is given by Eq.~\eqref{eq:ExpDef_CG} and the ``2 Digons'' subgroup is given by Eq.~\eqref{eq:ExpDef_CG2}.  If only a single time point is observed, these structures collapses to line segments.  The least face corresponds to the ``point'' topology.  The Hasse diagram illustrates the ordering square $\prec$ triangle $\prec$ line segment 1 $\prec$ point.  There is no relation between the triangle and digons subgroups or their collapsed line segments.  Note that the ordering is not graded; the square, triangle, and digons are each two-dimensional structures. }
\end{figure}

Being a poset, the diffeomorphism subgroups can be represented by a Hasse diagram, as we demonstrate for the exponential model from Eq.~\eqref{eq:ExpDef} in Figure~\ref{fig:HasseGroup}.  At the top of the diagram is the group preserving the square topology as in Figures~\ref{fig:Exps_1} and \ref{fig:Exps_2}.  Below this are the subgroups whose topologies are a triangle as in Figure~\ref{fig:Exps_CG_2} and Eq.~\eqref{eq:ExpDef_CG} and two digons as in Figure~\ref{fig:2Digons}.  If the observations are further coarsened so that only one time point is observed, these topologies collapse to line segments.  Finally, if nothing is observed we arrive at the least face, i.e., a single point.

Unlike the Hasse diagrams of section~\ref{sec:Hasse}, the Hasse diagram for the diffeomorphism subgroups is not graded.  In particular, note in Figure~\ref{fig:HasseGroup} that triangle $\prec$ square even though they are both two dimensional.  Similar to the Hasse diagrams in section~\ref{sec:Hasse}, the diffeomorphism subgroups will always include a minimal element that corresponds to the empty set (i.e., no observations) whose topology is a single point.  We speculate, that for most models there will also exist a unique ``maximal topology'' (i.e., greatest face).  The square is the unique maximal topology for the exponential model, and the pentahedron in Figure~\ref{fig:ESR_1} is the maximal topology for the enzyme-substrate model.

\section{Topological embedding defines stability of model classes} 
\label{sec:embedding}

\subsection{Model classes are ground states of statistical mechanics models}
\label{sec:groundstates}

In this section, we consider the topology of hierarchical models.  We introduce a concept called \emph{topological embedding}, i.e., how topologies of small models are embedded within larger ones.  We find that topological embeddings allows us to define the concept of stability for behavioral modes with respect to structural changes in the abstract model.  We motivate this definition in the context of log-linear statistical mechanics models, such as cluster expansions of binary alloys.  We therefore begin by discussing the topology for these models and its physical significance.

Consider a simple cluster expansion on a two-dimensional four by four square lattice with periodic boundary conditions.  Sites can be occupied by one of two atom types, i.e., a binary system.  The Hamiltonian for binary interactions up to fifth near-est neighbor takes the form
\begin{equation}
  \label{eq:2DSquareH}
  H = - \sum_{k=1}^5 \sum_{ d(i,j) = d_k} J_k s_i s_j,
\end{equation}
where the first sum indicates a sum over interactions and the second sum indicates a sum over sites such that the distance between sites $i$ and $j$ is equal to the $k^{th}$ nearest neighbor distance.  The random variables $s_i$ take on values $\pm 1$, indicating which atom type occupies the site, according to a Boltzmann distribution $P(\mathbf{s}) \propto E^{-H}$ where we have absorbed the temperature dependence into the definition of $J_k$.  This model has 16 binary random variables and therefore can exhibit $2^{16} = 65536$ distinct configurations.  Many of these states are symmetrically equivalent; only 432 configurations are crystallographically distinct.

\begin{figure}
\includegraphics[width=\linewidth]{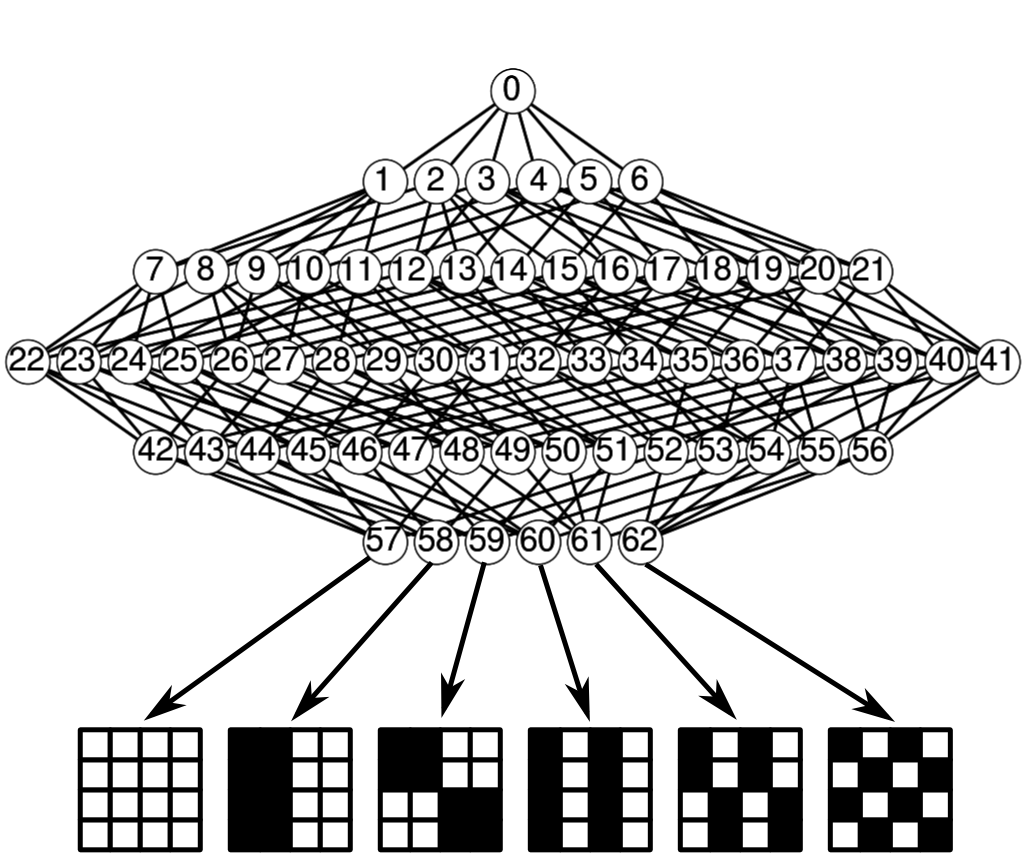}
\caption[]{\label{fig:IsingSquareTree} \textbf{Hasse diagram of a square-lattice, fifth-nearest neighbor model.}  The vertices of the model correspond to all of the ground state energy configurations for this model class.  Paths from the greatest face (labeled 0) to the vertices (labeled 57 - 62) correspond to systematic coarsening of available configurations.  For example, the path along nodes $0 \rightarrow 1 \rightarrow 7 \rightarrow 22 \rightarrow 42 \rightarrow 57$ systematically removes the high frequency configurations.}
\end{figure}

The topology of this model is summarized by the Hasse diagram in Fig.~\ref{fig:IsingSquareTree} (omitting the least face).  Although this model corresponds to a five-dimensional manifold (and is thus difficult to visualize), from the Hasse diagram it can be seen that the manifold has six hyper-corners corresponding to zero-dimensional limiting cases (six nodes on the bottom row).  These six points form the boundaries for 15 lines (second row from the bottom).  Continuing to work up the graph, these lines form the boundaries for 20 triangles, which form the boundaries for 15 tetrahedra, which form the boundaries for 6 first-order boundaries of the manifold.  These first order boundaries are each 4 dimensional objects bounded by five tetrahedra.  

As we have seen, the low dimensional elements of the Hasse diagram are models characterizing the behavioral models of the model which we now interpret.  Moving from the greatest face to a first order boundary (i.e.~any node from 1 to 6) corresponds to a limiting case in which $J_k \rightarrow \pm \infty$, i.e., a zero-temperature limit.  In these limits, many of the possible configurations are ``frozen out'' so that the models corresponding to the nodes 1 through 6 have between 21 and 65 structurally distinct configurations with nonzero probability (rather than the 432 in the complete model).  Although this is a low-temperature limit, the remaining configurations are not limited to the ground states.  Rather, the discarded configurations are those that are first to become irrelevant at low temperatures.

Moving down the Hasse diagram results in similar limits.  At each level, more and more configurations are removed.  Each level removes the next group of configurations least relevant at low temperatures.  The final result of this process is that only the ground states remain for models of dimension 0.  Each vertex corresponds to a model with only a ground state configuration.  These ground states are illustrated in Fig.~\ref{fig:IsingSquareTree}.

The process of systematically removing high-energy states from a model is reminiscent of a renormalization group procedure.  This analogy is reinforced by the observation that the sequence of approximations corresponding to the path labeled $0 \rightarrow 1 \rightarrow 7 \rightarrow 22 \rightarrow 42 \rightarrow 57$ removes configurations in order of their Fourier frequencies, from highest to lowest.  However, for many parameter values, high-frequency configurations do not correspond to high energy, the anti-ferromagnet being the seminal example.  For the anti-ferromagnet, the checkerboard configuration is a ground state (node 62).  

The topological structure of the model identifies the sequence of configurations least relevant and leads to a series of approximate effective models in which these configurations are removed.  The superficial similarity to a renormalization group procedure alluded to here will be explored in more detail elsewhere.

We now consider a similar model, a cluster expansion of a binary alloy on an FCC lattice and ask, what are the possible ground state configurations for models with different numbers of clusters?  This model takes the same basic form as before:
\begin{equation}
\label{eq:Hfcc}
H(\mathbf{s}) = \sum_i \Pi_i(\mathbf{s}) \theta_i,
\end{equation}
where $\mathbf{s}$ is the configurations of the two atom types on the lattice and $\Pi_i(\mathbf{s})$ are the energy contributions of a cluster expansion.  The parameters then are the contributions to the energy from nearest neighbor, next nearest neighbor interactions, etc., but can include also many body interactions.  For a real alloy, one expects that the true energy involves contributions from all of these terms.  However, including all order of the cluster expansion is both impractical and not theoretically enlightening.  In practice, the sum is truncated after a finite number of clusters.  How do the possible ground states depend on this truncation?

Although contributions to the total energy will be dominated by a few large terms in the sum (e.g., binary interactions of nearby neighbors) while other terms are less important (many body interactions of distant neighbors).  However, the question of which are the relevant parameters to include is more complicated than simply identifying which parameters are small.  In particular, some configurations may be unstable to small perturbations in other parameters.  Identifying an appropriate model involves identifying the appropriate parameter combinations that produce ground states that are stable with respect to the perturbation in the omitted parameters.

\subsection{Topological embedding identifies relevant and irrelevant parameters}
\label{sec:relevantparams}

\begin{figure}
\includegraphics[width=\linewidth]{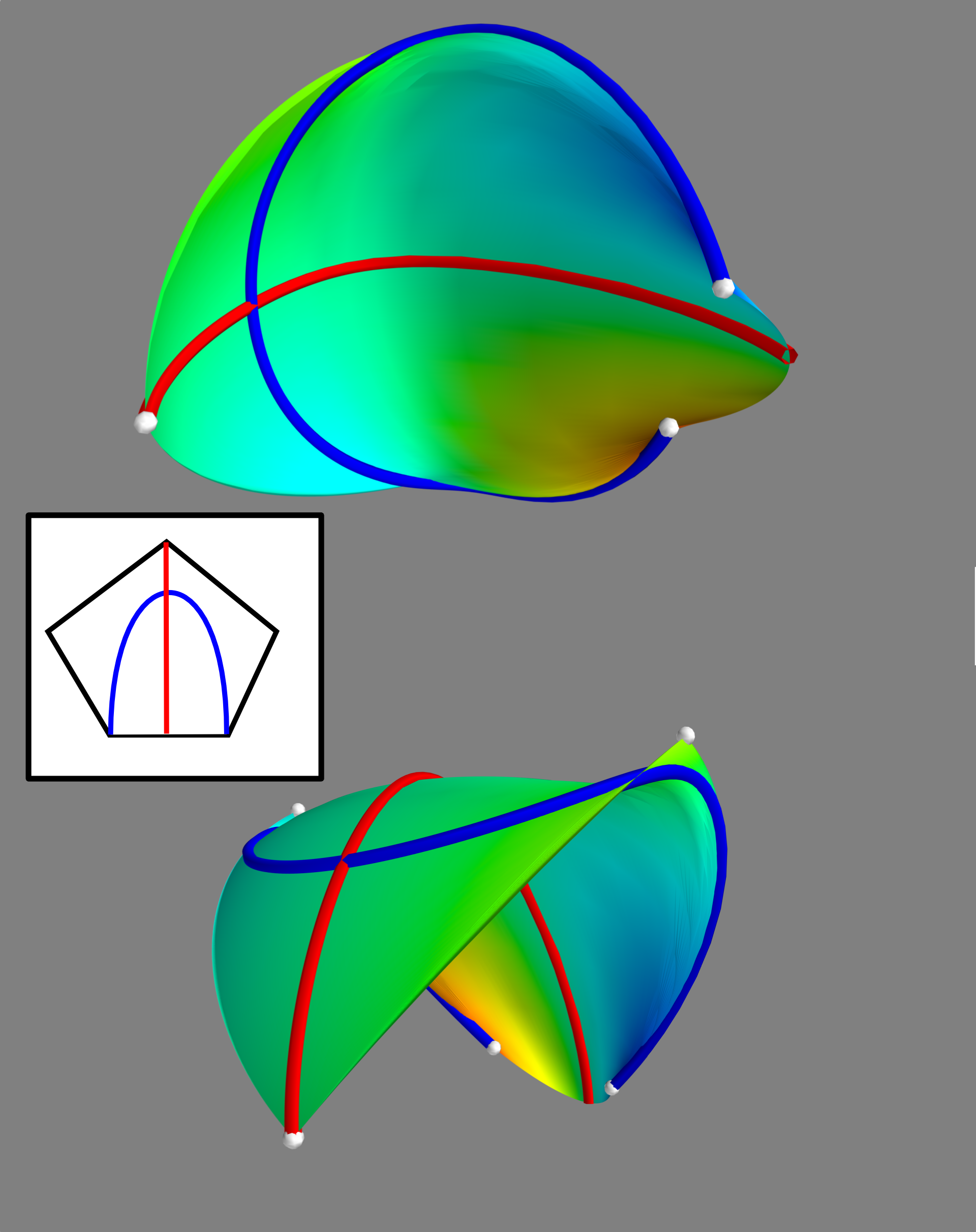}
\caption[]{\label{fig:FCC2} \textbf{Two parameter FCC lattice model.}  This model is topologically a pentagon.  Corners are indicated by white dots.  Two sub-manifolds (red and blue lines) correspond to the models in which one of the two parameters is fixed to zero.  The blue curve connects two ground states for the two parameter system, indicating that the ground states for this one parameter model are \emph{stable} to small perturbations in the second parameter.  The red curve, on the other hand, ends on an edge of the two parameter model, indicating that the corresponding ground state is \emph{unstable} to small perturbations in the second parameter.  The topological embedding of these models is summarized by the hierarchical graph in Figure~\ref{fig:FCC2_tree}. The cartoon inset summarizes the topological relationship among the three manifolds.}
\end{figure}

We now consider a sequence of models defined in Eq.~\eqref{eq:Hfcc} found by truncating the series at different numbers of parameters.  For a binary alloy on an FCC lattice, the manifold for a two parameter model with both nearest and second nearest neighbor two-body interactions is a pentagon, illustrated in Figure~\ref{fig:FCC2}.   We now consider the effect of truncating this this model to a one parameter model.  We do this by setting each of the two parameters to zero and varying the other parameter.  These one parameter models are illustrated by the red and blue curves in Figure~\ref{fig:FCC2}.

\begin{figure}
\includegraphics[width=\linewidth]{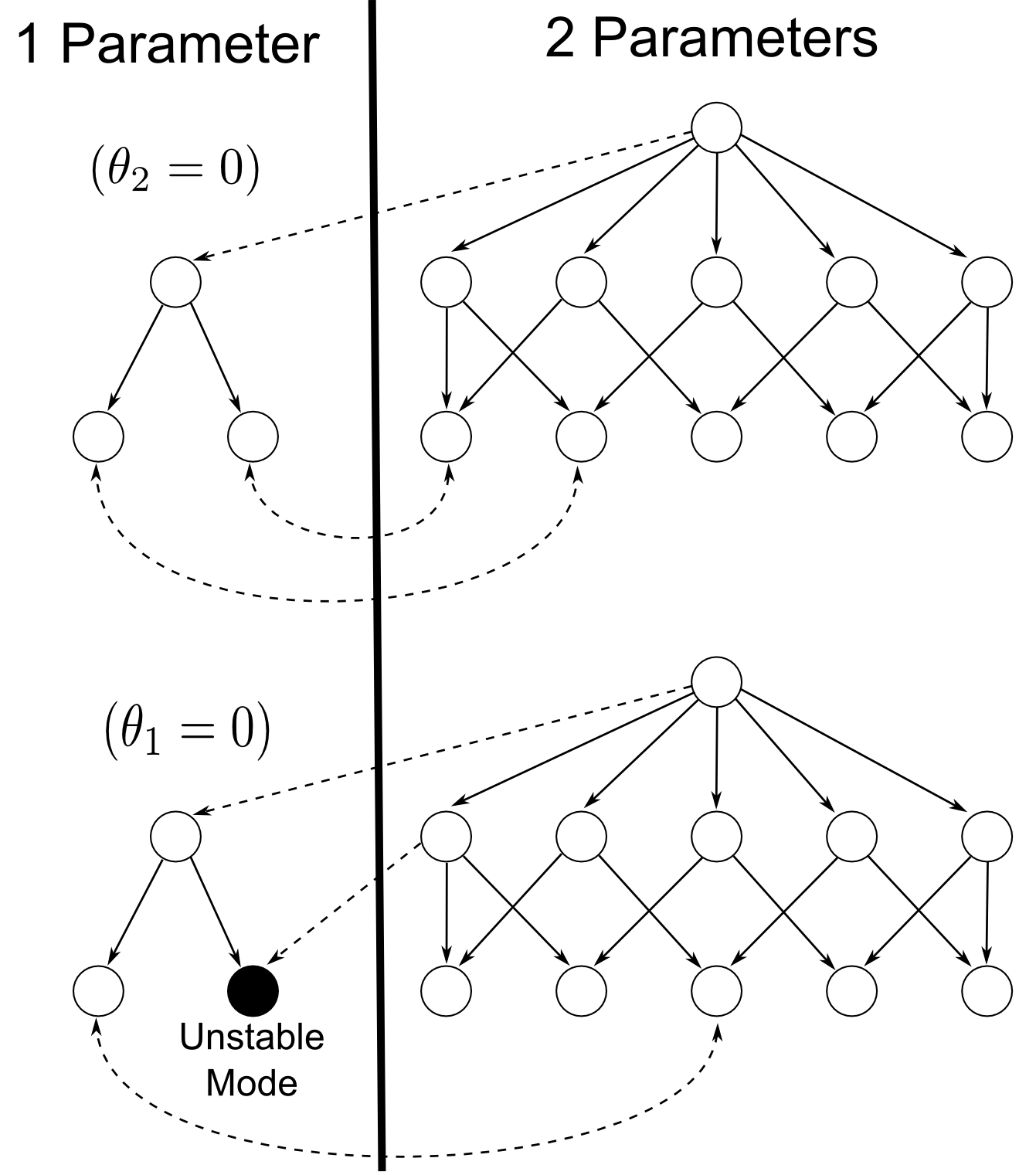}
\caption[]{\label{fig:FCC2_tree} \textbf{Topological embedding summarized by Hasse diagrams.}  The two parameter model visualized in Figure~\ref{fig:FCC2} is topologically a pentagon summarized by the Hasse diagram on the right.  The two models of one parameter (red and blue curves in Figure~\ref{fig:FCC2}) are line segments (Hasse diagrams on the left) that are embedded in the pentagon.  Embedding relationships are indicated by dashed lines across Hasse diagrams.  The ground states of the blue curve ($\theta_2 = 0$, above) are stable to small perturbations in $\theta_2$ and are also ground states of the two parameter model (above).  Dashed lines with two arrows indicate that nodes are equivalent.  One ground state of the red curve ($\theta_1 = 0$) is stable to small perturbations in $\theta_1$ and is likewise a ground state of the two parameter model.  The second ground state (colored black) is unstable because it is a subset of an edge of the two parameter model, indicated by a dashed line with one arrow.  This ground state will be unstable to small perturbations in $\theta_1$.}
\end{figure}

The corners of the two parameter model are illustrated by white dots in Figure~\ref{fig:FCC2}.  These dots represent the emergent model classes and behavioral models of the two parameter model (in this case the distinct ground states of the alloy).  Because the one parameter models are each included as special cases of the two parameter model, their topology is embedded within the topology of the two parameter model.  We summarize the relationship between the topologies of two parameter model and each of the one parameter models by the dashed lines in Figure~\ref{fig:FCC2_tree}.  For example, because each of the red and blue curves are subsets of the full two parameter model, we draw a dashed line from the greatest face of the two parameter graph to the greatest face of each of the one parameter  diagrams.  Furthermore, because both vertices of the blue curve correspond to corners of the full model, we connect them to the appropriate vertices of the two-dimensional diagram with dashed lines with two arrows to indicate equivalence (Fig.~\ref{fig:FCC2_tree}, top).  In contrast, only one vertex of the red curve is also a vertex of the two parameter model.  The other vertex intersects an edge of the two-parameter model.  We indicate that this vertex is a subset of an edge by a dashed line connecting the relevant nodes of the graph (Fig.~\ref{fig:FCC2_tree}, bottom).

To summarize, the vertices of the red and blue curves in Figure~\ref{fig:FCC2} represent the ground states for the one parameter models.  One of the ground states for the red curve is \emph{not} a ground state for the two parameter model and using this simple model would incorrectly predict the existence of a stable structure corresponding to this vertex.  The reason for this is not that the missing parameter is large in an actual alloy system, but that that configuration is unstable to arbitrarily small perturbations in this parameter.

This instability is manifest in the way the one parameter topology is embedded in that of the two parameter model.  Since the vertex of one parameter model lies on an edge and not a vertex of the two parameter model, the second parameter must be tuned to realize this behavior.

If a model class is unstable to the addition of a new parameter, we classify this parameter as relevant for modeling the behavior.  Similarly, if a model class is stable to the introduction of a new parameter, we classify the parameter as irrelevant.  Qualitatively, relevant parameters are those that must be tuned to realize a behavior.

The terms stable/unstable and relevant/irrelevant are used in analogy with similar definitions in RG analysis.  This present discussion has revolved around finding ground states for alloy models (i.e., phase transitions) in order to make the connection to these standard definitions more transparent.  An RG fixed point with no unstable direction (i.e., a sink) corresponds to a bulk phase\cite{goldenfeld1992lectures,zinn2007phase} because the system behavior in stable to variations in the parameters.  Near a fixed point, relevant parameters are those that need to be tuned to realize a phase transition or a critical point.  The language of information topology allows the same notions of stability of system behavior to extend to other diverse systems.  In this way, the question of what representation is appropriate for a particular physical system can be answered in a systematic way.

\section{Discussion}
\label{sec:discussion}

In this paper we have developed a mathematical language for exploring the informational relationships between models and observations.  A cornerstone of this formalism is the distinction between an abstract model and a statistical model.  The former refers to a collection of physical principles encoded in a parameterization, such as chemical reaction rates, and a set of rules for making quantifiable predictions, such as the law of mass action.  Application of these physical principles to specific experimental conditions leads to precise, quantifiable predictions that we refer to as the statistical model.  In other words, a statistical model is a realization of an abstract model for specific observations, so that there are many statistical models that share a common parameter space.

The distinguishability of predictions for different parameter values induces a metric on the parameter space.  Varying the observations changes the statistical  model, which in turn changes the geometric properties of the parameter space, such as distance and curvature.  However, if we consider the observations for which the parameterization remains structurally identifiable (at least in the local sense), then the resulting differential topology characterizes the abstract model.  We refer to the subsequent topological analysis as Information Topology.  

We have seen that for many models, topological properties can be visualized as a hierarchical graph known as a Hasse diagram.  A Hasse diagram naturally reveals the hierarchical relationship among the model's parameters. We identify the low-dimensional nodes of the Hasse diagram as emergent model classes that describe the behavioral modes the system.  The complete system behavior is understood as a combination of these modes.  Approximate effective models are systematically constructed from nodes of lower dimension in the Hasse diagram.  This procedure is made explicit by the manifold boundary approximation method\cite{transtrum2014model}. 



For hierarchical models, we have shown how the topology of small models is embedded in the larger model.  In this way, we characterize behavioral modes of the model as being either stable or unstable to variations in other parameter values.  This distinction leads naturally to the classification of parameters as either relevant or irrelevant depending on whether or not they need to be tuned to realize a behavior.  This classification is motivated by the similar classification originating in renormalization group analysis.

The framework presented in this paper provides new tools for understanding the relationship between abstract and statistical models and their manifest behaviors.  The problem of mathematical modeling in complex systems is an important one that spans many disciplines.  We anticipate that the concept of an information topology will be useful for studying systems in statistical mechanics, biology, chemistry, engineering, climate, and economics among others.  Application of the concepts in this paper may also be useful for problems related to experimental design, engineering and control, as well as providing a deeper understanding and explanation of the emergent physical principles that govern behavior in complex systems.

The authors thank Lei Huang, Jim Sethna, and Chris Myers for suggesting Figure~\ref{fig:LeiFig} and Kolten Barfuss and Alexander Shumway for helpful conversations.

\bibliographystyle{aps}
\bibliography{../../References/References}

\begin{thebibliography}{10}

\bibitem{wigner1995unreasonable}
E.~Wigner:
\newblock \emph{Wigner, EP op. cit}  (1995) 534

\bibitem{anderson1972more}
P.~W. Anderson, \emph{et~al.}:
\newblock \emph{Science} \textbf{177}  (1972) 393

\bibitem{machta2013parameter}
B.~B. Machta, R.~Chachra, M.~K. Transtrum, J.~P. Sethna:
\newblock \emph{Science} \textbf{342}  (2013) 604

\bibitem{rao1949distance}
C.~R. Rao:
\newblock \emph{Sankhya: The Indian Journal of Statistics} \textbf{9}  (1949)
  246

\bibitem{beale1960confidence}
E.~Beale:
\newblock \emph{Journal of the Royal Statistical Society. Series B
  (Methodological)}  (1960) 41

\bibitem{bates1980relative}
D.~M. Bates, D.~G. Watts:
\newblock \emph{Journal of the Royal Statistical Society. Series B
  (Methodological)}  (1980) 1

\bibitem{amari1985differential}
S.-i. Amari:
\newblock \emph{Differential-geometrical methods in statistics}:
\newblock Springer (1985)

\bibitem{amari1987differential}
S.-I. Amari, O.~E. Barndorff-Nielsen, R.~Kass, S.~Lauritzen, C.~Rao:
\newblock \emph{Lecture Notes-Monograph Series}  (1987) i

\bibitem{kass1989geometry}
R.~E. Kass:
\newblock \emph{Statistical Science}  (1989) 188

\bibitem{murray1993differential}
M.~K. Murray, J.~W. Rice:
\newblock \emph{Differential geometry and statistics}, vol.~48:
\newblock CRC Press (1993)

\bibitem{amari2007methods}
S.-i. Amari, H.~Nagaoka:
\newblock \emph{Methods of information geometry}, vol. 191:
\newblock American Mathematical Soc. (2007)

\bibitem{transtrum2010nonlinear}
M.~K. Transtrum, B.~B. Machta, J.~P. Sethna:
\newblock \emph{Physical review letters} \textbf{104}  (2010) 060201

\bibitem{transtrum2011geometry}
M.~K. Transtrum, B.~B. Machta, J.~P. Sethna:
\newblock \emph{Physical Review E} \textbf{83}  (2011) 036701

\bibitem{brown2003statistical}
K.~S. Brown, J.~P. Sethna:
\newblock \emph{Physical Review E} \textbf{68}  (2003) 021904

\bibitem{brown2004statistical}
K.~S. Brown, C.~C. Hill, G.~A. Calero, C.~R. Myers, K.~H. Lee, J.~P. Sethna,
  R.~A. Cerione:
\newblock \emph{Physical biology} \textbf{1}  (2004) 184

\bibitem{frederiksen2004bayesian}
S.~L. Frederiksen, K.~W. Jacobsen, K.~S. Brown, J.~P. Sethna:
\newblock \emph{Physical review letters} \textbf{93}  (2004) 165501

\bibitem{waterfall2006sloppy}
J.~J. Waterfall, F.~P. Casey, R.~N. Gutenkunst, K.~S. Brown, C.~R. Myers, P.~W.
  Brouwer, V.~Elser, J.~P. Sethna:
\newblock \emph{Physical review letters} \textbf{97}  (2006) 150601

\bibitem{gutenkunst2007universally}
R.~N. Gutenkunst, J.~J. Waterfall, F.~P. Casey, K.~S. Brown, C.~R. Myers, J.~P.
  Sethna:
\newblock \emph{PLoS computational biology} \textbf{3}  (2007) e189

\bibitem{casey2007optimal}
F.~P. Casey, D.~Baird, Q.~Feng, R.~N. Gutenkunst, J.~J. Waterfall, C.~R. Myers,
  K.~S. Brown, R.~A. Cerione, J.~P. Sethna:
\newblock \emph{IET systems biology} \textbf{1}  (2007) 190

\bibitem{daniels2008sloppiness}
B.~C. Daniels, Y.-J. Chen, J.~P. Sethna, R.~N. Gutenkunst, C.~R. Myers:
\newblock \emph{Current opinion in biotechnology} \textbf{19}  (2008) 389

\bibitem{transtrum2014model}
M.~K. Transtrum, P.~Qiu:
\newblock \emph{Physical Review Letters} \textbf{113}  (2014) 098701

\bibitem{goldenfeld1992lectures}
N.~Goldenfeld  (1992)

\bibitem{zinn2007phase}
J.~Zinn-Justin:
\newblock \emph{Phase transitions and renormalization group}:
\newblock Oxford University Press (2007)

\bibitem{eisenberg2013identifiability}
M.~C. Eisenberg, S.~L. Robertson, J.~H. Tien:
\newblock \emph{Journal of theoretical biology} \textbf{324}  (2013) 84

\bibitem{rothenberg1971identification}
T.~J. Rothenberg:
\newblock \emph{Econometrica: Journal of the Econometric Society}  (1971) 577

\bibitem{cobelli1980parameter}
C.~Cobelli, J.~J. Distefano~3rd:
\newblock \emph{American Journal of Physiology-Regulatory, Integrative and
  Comparative Physiology} \textbf{239}  (1980) R7

\bibitem{faller2003simulation}
D.~Faller, U.~Klingm{\"u}ller, J.~Timmer:
\newblock \emph{Simulation} \textbf{79}  (2003) 717

\bibitem{cho2003experimental}
K.-H. Cho, S.-Y. Shin, W.~Kolch, O.~Wolkenhauer:
\newblock \emph{Simulation} \textbf{79}  (2003) 726

\bibitem{balsa2008computational}
E.~Balsa-Canto, A.~A. Alonso, J.~R. Banga:
\newblock \emph{IET systems biology} \textbf{2}  (2008) 163

\bibitem{apgar2008stimulus}
J.~F. Apgar, J.~E. Toettcher, D.~Endy, F.~M. White, B.~Tidor:
\newblock \emph{PLoS computational biology} \textbf{4}  (2008) e30

\bibitem{apgar2010sloppy}
J.~F. Apgar, D.~K. Witmer, F.~M. White, B.~Tidor:
\newblock \emph{Molecular BioSystems} \textbf{6}  (2010) 1890

\bibitem{erguler2011practical}
K.~Erguler, M.~P. Stumpf:
\newblock \emph{Molecular BioSystems} \textbf{7}  (2011) 1593

\bibitem{chachra2011comment}
R.~Chachra, M.~K. Transtrum, J.~P. Sethna:
\newblock \emph{Molecular BioSystems} \textbf{7}  (2011) 2522

\bibitem{transtrum2012optimal}
M.~K. Transtrum, P.~Qiu:
\newblock \emph{BMC bioinformatics} \textbf{13}  (2012) 181

\bibitem{transtrum2016manifold}
M.~K. Transtrum:
\newblock \emph{arXiv preprint arXiv:1605.08705}  (2016)

\bibitem{grunbaum1967convex}
B.~Grunbaum, V.~Klee, M.~A. Perles, G.~C. Shephard:
\newblock \emph{Convex polytopes}:
\newblock Springer (1967)

\bibitem{ziegler1995lectures}
G.~M. Ziegler:
\newblock \emph{Lectures on polytopes}, vol. 152:
\newblock Springer Science \& Business Media (1995)

\bibitem{brondsted2012introduction}
A.~Brondsted:
\newblock \emph{An introduction to convex polytopes}, vol.~90:
\newblock Springer Science \& Business Media (2012)

\end{thebibliography}

\section*{Supplemental Information}

\subsection{Visualization Methods}

Most examples throughout the text were chosen to consist of either two or three parameters, corresponding to manifolds of two or three dimensions, so that their topological structure could be more easily visualized.  However, these manifolds are embedded in a data space of often much higher dimension.  We therefore project this high-dimensional embedding space into a three dimensional subspace in order to generate the figures found in the main text.

In order to generate low-dimensional projections, we first constructed a grid of the parameter space and evaluated the model prediction at each point.  The corresponding grid of model predictions correspond to a grid of vectors in the high-dimensional embedding space.  We then performed a principle component analysis of this grid of prediction vectors and projected the grid onto each of the principle components.  We then created the visualization by truncating all but the first three principle components.  For two-dimensional manifolds, we have colored the manifold according to the fourth principle component, effectively visualizing a four-dimensional embedding.  For three-dimensional manifolds, i.e., the enzyme-substrate model, we have colored each boundary of the volume a solid color to help illustrate the topological structure.

\subsection{Embedding space for non least-squares models}

We have used the Fisher Information Matrix (FIM) as the Riemannian metric for measuring distances between models.  When the model in question is a least-squares model, then there is a natural Euclidean metric in behavior space corresponding to the sum of squares defined by the $\chi^2$ cost function.  This Euclidean metric induces a non-euclidean metric on the embedded model that is equivalent to the Fisher Information\cite{rao1949distance,beale1960confidence,bates1980relative,amari1985differential,amari1987differential,kass1989geometry,murray1993differential,amari2007methods,transtrum2010nonlinear,transtrum2011geometry}.

A similar Euclidean embedding space can be found for an arbitrary probabilistic model $P(\xi, \theta)$.  We derive this metric for the case of discrete probability distributions, but the result also holds for probability densities.  This embedding space can be found by constructing the variable $z_i(\theta) = 2 \sqrt{P_i(\theta)}$ where the index $i$ labels each configuration of the random variable $\xi$.  Assuming a Euclidean metric on the vector $z$, gives 
\begin{equation}
ds^2 = \sum_i dz_i^2 = \sum_i \frac{dP^2_i}{P_i} =\left\langle \frac{dP^2}{P^2} \right\rangle,
\end{equation}
where in the last term we have written the sum as an expectation value.  Note that 
\begin{equation}
\frac{dP}{P} = d \log P = \sum_\mu \frac{\partial P}{\partial \theta_\mu} d \theta^\mu,
\end{equation}
from which it follows that
\begin{equation}
ds^2 = \sum_{\mu\nu} \left\langle \frac{\partial \log P}{\partial \theta_\mu}  \frac{\partial \log P}{\partial \theta_\nu} \right\rangle d \theta^\mu d \theta^\nu,
\end{equation}
so that the induced metric on the manifold is 
\begin{equation}
g_{\mu\nu} = \left\langle \frac{\partial \log P }{ \partial \theta_\mu } \frac{\partial \log P }{ \partial \theta_\nu }\right\rangle,
\end{equation} 
which is the Fisher Information Metric.

Notice that normalization requires that $\sum_i z_i^2 = 4 \sum_i P_i^2 = 4$, so that the variables $z_i$ correspond to a hyper-sphere embedded in a Euclidean data space.  The model manifold is then embedded in this hyper-sphere.  We interpret this Euclidean, hyper-sphere embedding in the same way as the Euclidean embedding space for least squares models, i.e., as a data or behavior space.  

\subsection{Topology and alternate metrics}

In the main text of the paper, we have used the Fisher Information Matrix (FIM) as the metric that defines relative distances between different statistical models that are realized from the same abstract model.  Since an abstract model is characterized by its parameterization, we found that changing the observations lead to different metrics on the same parameter space.  Although the FIM has many desirable statistical properties, it is not the unique measure of statistical distance between probability distributions.  Other notable examples include the Hellinger distance, total variation distance, the Levy-Prokhorov metric, Bhattacharyya distance, earth mover's distance (also known as the Wasserstein or Kantorovich metric) and the energy distance.  Furthermore, there are many other distance measures that are applicable to specific types of models and not probability distributions generally, such as the $H_\infty$ norm in control theory applicable to dynamical systems.  (While all of these metrics are valid measures of statistical distance, they are not all \emph{Riemannian metrics}, i.e., they do not define an inner product.  Indeed, one of the advantages of the FIM is that it corresponds to a Riemannian metric, thus allowing the use of differential geometry.)

Changing metrics among the many possible choices is in many ways analogous to changing observations.  The net effect is to change the meaning of distance for parameter values in the model.  As such, the results of this paper can be applied to alternative metrics.

Furthermore, there are many measures of \emph{statistical divergence} that, while not properly metrics (they may not satisfy either symmetry or the triangle inequality), are often interpreted as a type of distance measure.  A class of such divergences are the f-divergences, among which the Kullback-Leibler divergence is a well-known example.  (Interestingly, the FIM corresponds to the f-divergence for infinitesimally separated distributions.)  Other classes of divergences include the M-divergences and S-divergences.  Specific examples include Renyi's divergence and the Jensen-Shannon divergence.  Although not properly metrics, statistical divergences do define a topological space.  Consquently, the topological properties of abstract model can realized using a divergence rather than a proper metric.

\end{document}